\documentclass[oneside,14pt]{book}
\usepackage{amsmath,amssymb,amsfonts,bm,marginnote,cite,graphicx}
\usepackage[colorlinks=true, pdfstartview=FitV, linkcolor=black, citecolor=black, urlcolor=black]{hyperref}
\usepackage{quotchap}
\usepackage{emptypage}
\usepackage[usenames,dvipsnames]{xcolor}
\usepackage{pdfpages}

\newcommand{\be}{\begin{equation}}
\newcommand{\ee}{\end{equation}}
\newcommand{\beq}{\begin{eqnarray}}
\newcommand{\eeq}{\end{eqnarray}}
\newcommand{\bea}{\begin{eqnarray}}
\newcommand{\eea}{\end{eqnarray}}
\newcommand{\beqn}{\begin{eqnarray}}
\newcommand{\eeqn}{\end{eqnarray}}

\newcommand{\X}{\mathbb{X}}

\newcommand{\Am}{\mathbb{A}}

\def\pa{\partial}
\def\om{\omega}
\newcommand{\rd}{\mathrm{d}}

\def\S{\mathbb{S}}
\def\s{\sigma}

\def\P{{\cal{P}}}
\def\Q{{\mathbb{Q}}}

\begin{document}

\includepdf[pages=-]{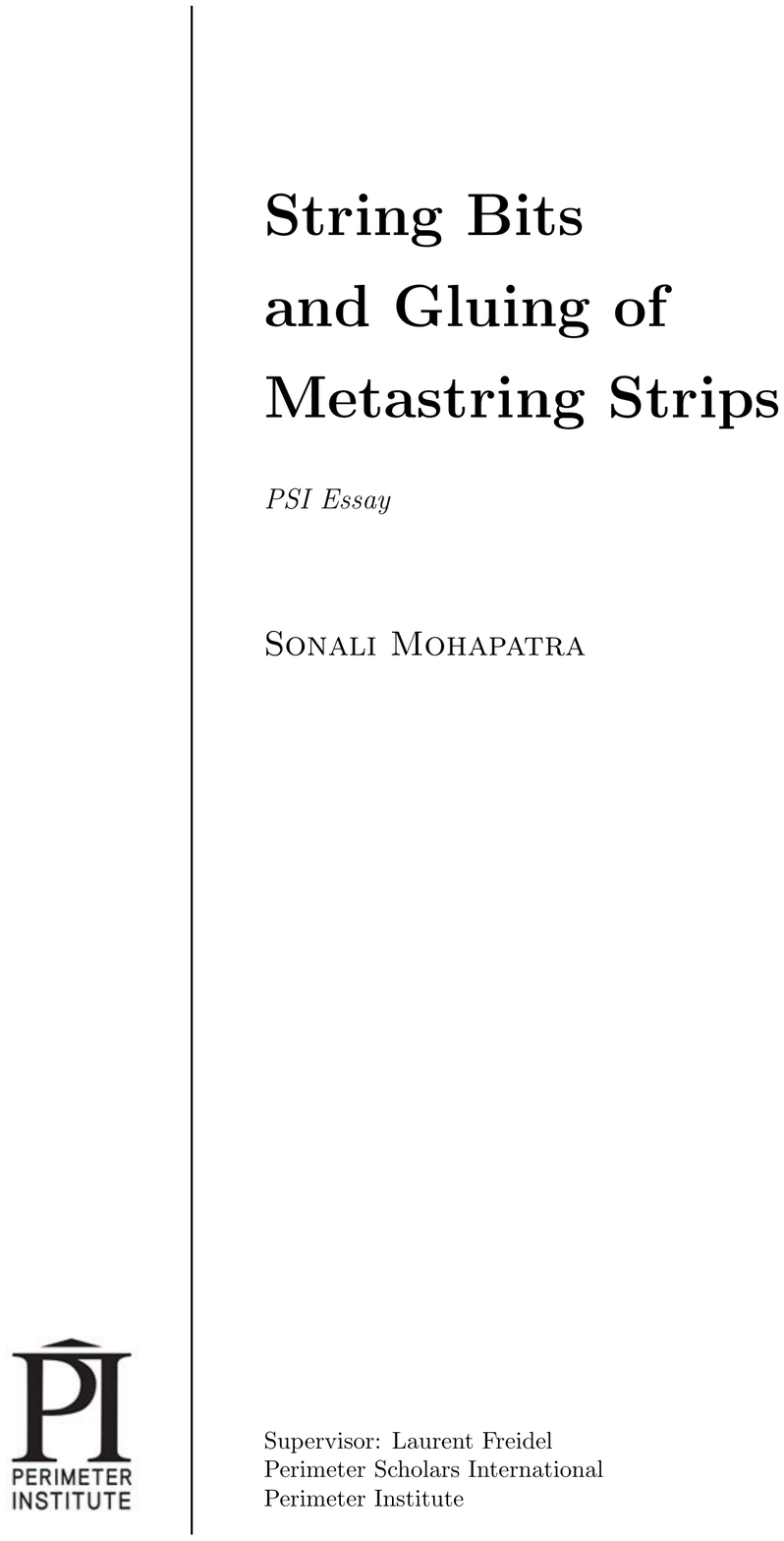}
\begin{titlepage}

\clearpage
\vspace*{\fill}
\begin{center}
\begin{minipage}{.6\textwidth}
\begin{center}
This essay has been submitted towards the completion of \\ 
Perimeters Scholars International (PSI) program at \\
Perimeter Institute, \\
affiliated to the \\
University of Waterloo, Canada \\
under the supervision of \\
Dr Laurent Freidel.\\
\hspace{2mm}

\textit{Class of 2015.}
\end{center}

\end{minipage}
\end{center}
\vfill 
\clearpage

\end{titlepage}

\begin{center}
\vspace{15mm}

\includegraphics[scale=1]{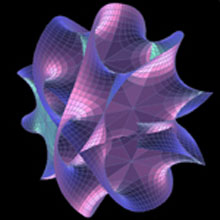} \\

\vspace{15mm}

``According to String Theory, \\
what appears to be empty space is actually \\
a tumultuous ocean of strings vibrating at the precise frequencies \\
that create the 4 dimensions \\
you and I call \\
height, width, depth and time."\\
\vspace{10mm}
-Roy H. Williams.

\end{center}

\chapter*{Abstract}
\addcontentsline{toc}{chapter}{Abstract}

In this essay, we review the meta-string formulation proposed by Freidel, Leigh, Minic in a recent paper. Our work focuses on the construction of a closed-string world-sheet from gluing of Nakamura strips. We review the symplectic current formulation for determining the gluing condition for a single strip. We then study the two-strip scenario in a new notation to rigorously derive the boundary equations of motion and finally generalize our result to N -strips. We find the conjugate momenta to the strip separation variable and relate it to the midpoint velocity of a strip. We see the natural evolution of meta-D-branes in meta- string theory from the strip picture. Finally, we motivate and show the connection of the strip picture to the string-bit picture put forward by Klebanov and Susskind and conjecture the relation to the Chan-Paton factors.

\chapter*{}
\addcontentsline{toc}{chapter}{Dedication 1}
\begin{center}
This thesis is dedicated to my grandfather (Aja).\\
Enthusiastic, supportive, science lover, \\
mathematician, guide and friend...\\
You will forever stay in our hearts and soul...\\
\end{center}
\vspace{15pt}

\begin{center}
For you, \\
\textbf{Aja, the best},\\

\vspace{5pt}

``Timeless you remain\\
 Sparkling you be. \\
 In the endless river, \\
 rescued by thee. \\
 One paradise lost,\\
 another found..\\
Timeless we are, \\
entwined and bound...''

\end{center}

\chapter*{}
\addcontentsline{toc}{chapter}{Dedication 2}

\begin{center}To Bou and Nana. \\
For always been my great wall at times of need and my well of happiness when I am thirsty.\end{center}
\vspace{15pt}
\begin{center}And when the sky is dark,\\
and the lights are out,\\
You taught me how to find my way home, \\
even when I am blind-folded.
\end{center}

\chapter*{Acknowledgements}
\addcontentsline{toc}{chapter}{Acknowledgements}

At the end of an illuminating year, I convey my sincere thanks to my supervisor, Dr Laurent Freidel, for the highly motivational discussion sessions, for putting up with my ramblings, joining in the philosophizing and for pushing me evermore on the path of curiosity. I am thankful to Perimeter Scholars International, the PSI fellows as well as the administration for making this work possible. I would like to thank my mentor Tibra Ali, for his helpful conversations and advice. To my fellow night-owls at the George Leibbrandt Library and the Green Room, I convey my happiness and thanks for being amazing friends, for infinitely long pointless rambling sessions, for sharing good music and for helping with many technical details. I am also deeply thankful to my best friend, Saurav Dutta for encouraging me at my lows and sharing my highs. I would also like to thank my relatives for being my home away from home. Last but not the least, I am deeply grateful to my family for letting me chase my dreams, bearing with my eccentricities and fuelling my curiosities. This work would not have been possible without their support.

\begin{small}
\tableofcontents
\end{small}

\begin{savequote}[45mm]
Everything that exists, exists somewhere and nothing happens that does not happen at some time.
\qauthor{Lee Smolin, Three Roads to Quantum Gravity}

I, at any rate, am convinced that he does not throw dice!
\qauthor{Albert Einstein}
\end{savequote}

\chapter{Introduction and Motivation}

Since the discovery of General Relativity in 1915 followed by Quantum Mechanics in 1926, our perception of space-time has radically changed. General Relativity came with the genius of Einstein that gravity must be the geometry of space-time and has soon become the most successful theory of gravity we have presently. In the mean-time, not to be left behind, Quantum theory has made rapid strides and has taken over as the framework for the other three forces, namely, weak, strong and electromagnetism. This surely led to the logical conclusion that quantum theory must lead to modifications in GR as well leading to a quantum theory of gravity.

\section{The Problem of Quantum Gravity}

But the search for this seemingly simple task has been the most elusive with gravity proving to be the most flighty of temptresses. Infinities here, infinities there and infinities everywhere has led to no consensus of how best to proceed with the job of reconciling quantum theory and GR. Moreover, the fact that we presently have no way of testing the theories we have at hand, be it string theory, LQG, causal sets, causal dynamical triangulations or quantum graffiti leave us helpless, grappling for better theories.``The problem of quantum gravity is simply the fact that the current theories are not capable of describing the quantum behaviour of the gravitational field [1].'' There seems to be an inherent incompatibility between the two theories. But ``Contradiction between empirically successful theories is not a curse: it is a terrific opportunity! [1]'', and in our opinion, string theory provides a handsome playground of opportunities.

\section{String Theory and Space for Modification}

Looking at the tenets of quantum mechanics, the duality of space-time and momentum space strikes us as unique. Position and Momentum, which are conjugates of each other are treated on an equal footing without any structural differences. The simplest example where this symmetry is fully manifest is the harmonic oscillator. In Einstein's GR, though, this symmetry is broken and a structural dichotomy is introduced. Space-time becomes curved whereas the momentum space is taken to be the tangent space to the curved space-time and is flat. Is it possible that the curving of momentum space might lead to understanding new physics at the planck length or maybe provide a natural way to unify both QM and GR? It seems reasonable to assume so.

String theory has emerged as the promise of unification of quantum gravity with other known theories and as a umbrella theory of everything. In [2], it has been argued that string theory provides a handsome
framework to curve the momentum space as well and remove this dichotomy. But in order to work with both space-time and momentum space on an equal footing, we must find a framework to unify them in the phase-space and deal with generalised phase-space variables. This reformulation of string theory in the phase-space is known as meta-string theory [2]. In order to build a clear understanding, let us list the salient points of the usual formulation of string theory which meta-string theory proposes to modify or relax:

\begin{enumerate}
\item A fixed background space-time target space. (The string is assumed to be propagating in space-time).
\item This implies the existence of the closed string boundary conditions, $\pa_{\s}X^{\mu}(\tau,l) = \pa_{\s}X^{\mu}(\tau,0)$ and
$X^{\mu}(\tau, l) = X^{\mu}(\tau, 0)$
\item The left movers and the right movers have the same centre of mass momentum $p^{\mu}$.
\end{enumerate}

\section{String Theory in Phase Space: the Meta-String}

Metastring theoy makes an interesting departure from the usual formulation of string theory by relaxing the assumption that the string is propagating on a fixed back-ground space-time. Discarding this notion of a fixed background space-time leads us to relax the requirement for the left movers and the right movers to have the same momentum allowing us to make our theory chiral. In this formulation, space-time and momentum space both emerge as lagrangian submanifolds of the symplectic target space. The target space, which in this formulation is the phase-space, seems to be a natural background for the meta-string to propagate in.

Discarding the notion of a fixed background space-time also makes us look deeper into the boundary condition, namely $ \pa_{\s} X^{\mu}(\tau, l) =\pa_{\s} X^{\mu}(\tau, 0)$. Now, this boundary condition does not imply $X^{\mu}(\tau, l) = X^{\mu}(\tau, 0) $ but rather more generally, $X^{\mu}(\tau, l) = X^{\mu}(\tau, 0) + \delta$  which introduces a monodromy on X. But having a monodromy on X would imply that we would not have a closed string worldsheet apriori! This will prove us a deterrent since, naively thinking, string theory predicts the existence of the graviton from the modes of the closed string propagation. Thus, we need to examine closely whether closed string world-sheets can exist in the meta-string formalism inspite of the presence of monodromies.

One-way of proceeding is provided by [2] in which the closed string world-sheet is cut open at a single value of  $\s$ and the monodromies are imposed. Now, thinking in terms of symplectic geometry, the strip can be glued back together at the cut only if the symplectic flux across the cut is conserved. Thus, we write the variation of the action and write the expression for the symplectic structure and the symplectic flux across the cut. Following this line of reasoning, we look at how we might form various closed string interaction pictures in meta-string theory. For example, a single strip such as in Figure 2.1 can be glued back together to form the simple closed string world-sheet cylinder. However, what about the string interaction pants diagram such as Figure 3.1? This would require us to glue a minimum of 2 strips together in a specific way. Similarly, it is easy to see that we would be able to decompose any string interaction diagram into a number of (nakamura) strips and would be able to glue them back together if the symplectic flux across the cut is continuous in the presence of the meta-string monodromies.

The first step in this path would be to examine whether the simple closed string world-sheet can be decomposed into a multi-strip picture and from there, proceed to the case with one interaction point (Figure 3.1), two interaction points and so on.. It is imperative we remember that in the meta-string description, the world-sheet is propagating in the phase-space rather than the space-time. This strip decomposition picture which is explored in detail in this essay, also brings to mind the string-bit picture as proposed by Klebanov and Susskind in [3] and [4]. The string-bits discovered by Klebanov and Susskind are the units of the quantum string which when glued together have the property of forming a continuous world-sheet while experiencing a discrete space-time, the dichotomy between the continuum world sheet and the discrete target being due to the conformal invariance of the string in the light cone gauge. This seems quite bizarre and the weirdness is quite clear if we put it poetically as Freidel: ``The continuous limit seems to be discrete!'' In the case of the meta-string, in the present semi-classical picture described in this essay and in [2], the meta-string strips seem to be the analog of the string bits.

Furthermore, we will also try to make connections to the Chan-Paton factors in Matrix models of string theory, since we will see that the addition of boundary terms to the action will lead us to write the amplitude of the worldsheet as a product over the strip amplitudes. A pleasant surprise comes from examining the boundary terms closely as they are the terms with the Dirichlet boundary conditions which brings about the evolution of D-branes in meta-string theory quite naturally from the strip picture.
In the next chapter, we will review certain parts of meta-string theory, such as the construction of the action as well as the symplectic analysis of the boundary. Chapters 3 and 4 deal with the strip-picture or \textit{``stripology''} in detail.

\begin{savequote}[45mm]
We know a lot of things, what we don’t know is even more.
\qauthor{Edward Witten}
\end{savequote}

\chapter{Construction of the Metastring Action}

In this chapter, we are going to review the formulation of the meta-string theory in some detail following the paper by Freidel et al [2].To formulate the metastring in the phase space of string theory, we start with the usual Polyakov action:

\beq
S_{P}(X) \equiv \frac1{4\pi\alpha '} \int_{\Sigma} h_{\mu\nu} (\star \rd X^{\mu} \wedge  \rd X^{\nu}),
\eeq

where $\Sigma$ is the 2-dimensional string worldsheet with local coordinates $(\tau,\s)$ and metric $ds^2 \equiv  -(d\tau)^2 + (d\s)^2$, $\star$ is the hodge dual on the world sheet such that $\star d\tau = d\s$ and $\star ^2 = 1$, d is the exterior derivative on the worldsheet and $X^{\mu}$ are local coordinates on a D-dimensional target space M with Minkowski metric $h_{\mu\nu}$, which we interpret as a (flat) spacetime. Now in order to make the action single-valued, we need to impose the condition $\pa_{\s}X$ is periodic. But this does not imply that X is periodic, rather X is quasi-periodic. Thus, we write,

\be X^{\mu}(\s + 2\pi,\tau) = X^{\mu}(\s,\tau) + \delta^{\mu}, \ee

where $\delta^{\mu}$ is the monodromy. Introducing an auxiliary one form $\bm{P}_{\mu}$ into the action, a momentum scale
$\varepsilon$ and a length scale $\lambda$, the Polyakov action becomes,

\be
\hat{S}(X, \bm{P}) \equiv \frac1{2\pi} \int_{\Sigma} \left( \frac{1}{\lambda \epsilon} \bm{P}_\mu \wedge\rd X^\mu+ \frac{1}{2\epsilon^2} h_{\mu\nu} (*\bm{P}_\mu \wedge \delta\bm{P}_\nu) \right),
\ee 

Varying the action with respect to $\bm{P}$, we get,

\be \bm{P}_{\mu} = \frac{\epsilon}{\lambda} h_{\mu\nu} dX^{\nu} \ee

and plug this back in to get the total action, $S_{T}$ to be

\be \hat{S}_{T}(\bm{P}) = -\frac1{4\pi(\epsilon)^{2}}\int_{\Sigma} h^{\mu \nu} (\star \bm{P}_{\mu} \wedge \bm{P}_{\nu}). \ee

Now, integrating out X, we see that $d\bm{P}_{\mu} = 0,$ i.e, $\bm{P}$ is closed. Thus, locally we can write
\be \bm{P}_{\mu} \equiv \rd Y_{\mu} \ee

where Y is a 0-form with dimensions of momentum. Now decomposing $\bm{P}$ as follows:

\be
\bm{P}_{\mu} = P_{\mu} \rd \sigma +  Q_{\mu}\rd \tau,
\ee

we plug it in and our action looks like

\be
\hat{S}_{T} = \frac{1}{4\pi\lambda \varepsilon}\int \left( P_{\mu} \pa_{\tau}{X}^{\mu} -  Q_{\mu} \pa_{\sigma}X^{\mu} + \frac{\lambda}{2\epsilon} (Q_{\mu}Q^{\mu}- P_{\mu}P^{\mu} )\right).
\ee

Integrating out Q by inserting its equation of motion, we find that Y is also quasi periodic, 

\be Y_{\mu}(\s + 2\pi,\tau) = Y_{\mu}(\s,\tau) + 2\pi p_{\mu} \ee

where $p_{\mu}$ is the string momentum. Now unifying  $X$ and $Y$ into a dimensionless coordinate $\X$ on phase space $\P$, we write

\be\label{XY}
	\X^{A} \equiv \left( \begin{array}{c} {X^{\mu}}/\lambda \\  Y_{\mu}/\epsilon\end{array} \right).
	\ee
	
where $\lambda$ and $\epsilon$ are the parameters with the dimension of length and momentum respectively making $X$ and $Y$ dimensionless. Now we need to make sure we have a way to project this action out onto $X$ (space-time) and $Y$ (momentum-space).

\section{Relative Locality}

In order to write this action in terms of phase space coordinates (equation 2.10), we need to introduce the structures, $\omega$, $\eta$ and H which are the symplectic structure on the phase space, the polarization metric, and the quantum or the generalized metric as follows:

\be\label{etaH0}
	\eta_{AB}^0 \equiv \left( \begin{array}{cc} 0 & \delta \\ \delta^{T}& 0  \end{array} \right),\quad
H_{AB}^0 \equiv  \left( \begin{array}{cc} h & 0 \\ 0 &  h^{-1}  \end{array} \right),
\quad \om_{AB}^0 = \left( \begin{array}{cc} 0 & \delta \\ -\delta^{T}& 0  \end{array} \right) ,
\ee

where $\delta^{\mu}_{\nu}$ is the d-dimensional identity matrix and $h_{\mu\nu}$ is the d-dimensional Lorentzian metric, $T$ denoting transpose. The two-form $\omega$ expresses the fact that $\P$ is a symplectic manifold and $(\eta+\omega)$ and $(\eta-\omega)$ act as projectors to the space-time and momentum space lagrangian submanifolds, respectively. We might worry about the fact that writing the action in such a double-field theoretical manner, doubles our dimension, thus in turn doubling our degrees of freedom. Our worries can be put to rest by making the theory chiral by introducing the Chiral metric $J$:

\be\label{Jleft} J|\text{leftmovers}\rangle = +1|\text{leftmovers}\rangle \ee

and 

\be\label{Jright} J|\text{rightmovers}\rangle = -1|\text{rightmovers}\rangle \ee

Moreover, the P-metric and the Q-metric are related in the present context through J as follows:

\be
J_0\equiv (\eta^{0})^{-1}H^0,
\ee

we see that $ J_{0}$ is an involutive transformation preserving $\eta^{0}$, that is,

\be
J^{2}_0=1,\quad \mathrm{ and} \quad  J^{T}_0\eta^0 J_0=\eta^0
\label{Jprops}
\ee

$(\eta^{0},J^{0})$ defines a \textit{chiral structure} on the phase space P.

Chirality is okay to impose, since in the usual formulation of string theory, there is no reason apriori from the CFT perspective as to why the left movers and the right movers have to have the same string momenta. This imposition comes from a assumption of locality of the target space which is space-time. This is another assumption we must relax in order to make the theory inherently quantum. We allow both $\eta$ and H to be fully dynamical. This is known as the theory of relative locality where each probe sees a different space-time. Interactions between various probes help us to agree upon a specific space-time. Thus $\eta$ = flat is a condition for absolute locality, where we recover the usual space-time concept as we know it. Using these definitions 2.11, the action is written as a $\s$-model on $\P$:

\be\label{flatTseytlin}
S_T= \frac1{4\pi}\int  \Big( \partial_{\tau}{\X}^{A} (\eta_{AB}^0 + \omega_{AB}^0)\partial_{\sigma}\X^{B} -   \partial_{\sigma}\X^{A} H_{AB}^0\partial_{\sigma}\X^{B} \Big).
\ee

We now call this the metastring action and will focus on this meta-string action for the rest of this book.

\section{T-Duality in Metastring Theory}

In the normal world sheet picture of conventional string theory, T-duality is realized as the symmetry of the world-sheet action where we exchange $\s$ and $\tau$. But as is manifest from the metastring action, we see that this is not the case. Instead, we have another symmetry in the action namely the map:

\be \X \mapsto J(\X) \ee

This is also one reason for demanding $J^2 = 1$. In fact, it turns out we need not fear as T-duality is not lost. It manifests as a symmetry of the target space which is the phase space in the new formulation.

\section{Nakamura Strips and Lorentzian Worldsheets}

In section (2), we reviewed the derivation of the metastring action which is the string action in the phase- space. We also saw that for the Polyakov action to be single valued we need to impose the condition of that $\pa_{\s}\X$ is periodic. But this implies that $\X$ itself might not be periodic and might include monodromies of the type:

\be \X(\s + 2\pi, \tau ) = \X(\s, \tau ) + \Delta \ee

where $\Delta$ is the monodromy of $\X$ and $\X$ is the phase space double variable. This is where it gets tricky. If we do not demand $\Delta$ to vanish, then there is no apriori space-time interpretation of the target space and no apriori closed string propagation cylinder in a background space-time. Since, the intent finally is to generalize T-duality in a curved background, we do not initially restrict the monodromies to have a space-time interpretation, instead we try to find the conditions they have to satisfy.

We can find from [2] that the Tseytlin action (equation 2.16) is not Lorentz invariant. But nevertheless, we would expect the full quantum theory to be Lorentz invariant, at least for the flat background. (In the action, this condition of having a flat background is imposed by choosing $\eta$, $\omega$ and H to be all constant.) The construction of the moduli space of Lorentzian worldsheet can be found in [2]. The lesson that we take away from [2] is that the construction with some minor modifications as shown in [2] again is covariant and modular invariant and thus, can be applied to arbitrary conformal field theories. By a powerful result of Giddings and Wolpert, we see that it is possible to decompose each general lorentzian worldsheet into a collection of flat strips, each of which can be coordinatized by locally flat coordinates. Each such decomposition gives rise to what is called as a Nakamura graph $N$. Moreover, it is given by the GWKN theorem that slicing the world-sheet into Nakamura strips is invariant under large diffeomorphisms. Moreover as seen in [2], we also conclude that we can consider the chiral phase space action and preserve modular invariance. Since we are interested in seeing whether metastring theory possesses the closed string propagation cylinder, we will begin a study of how to sew the strips back together to form a closed cylinder in the next section.

\section{Strip Decomposition and Gluing}

In order to examine whether a closed string interpretation exists in the metastring formalism, we review [2] as they consider a cylindrical worldsheet geometrical structure $\Sigma$. This has been cut open along $\s = (0, 2\pi)$. Thus, we are considering a single strip as in Figure 2.1. As we saw earlier, the strip action can be written as follows:

\beq\label{actionpatch}
S_T=\frac1{2} \int_{\tau_o}^{\tau_i}\rd \tau \int_{0}^{2\pi} \rd \s\left[  \pa_{\tau}\X^{A} (\eta_{AB}+ \omega_{AB})\pa_{\s}\X^{B} -  
  \pa_{\s}\X^{A} H_{AB} \pa_{\s}\X^{B}\right] ,
\eeq

where we have restricted to $S$, coordinatized by $\tau \in [\tau_{i},\tau_{o}], \s \in [0,2\pi)$. The fields $\X$ are assumed to be quasi-periodic $\X(\tau,\s+2\pi) = \X(\tau,\s)+\Delta(\tau)$, with $\pa_{\s}\Delta = 0$. Note that this implies that $\pa_{\s}\X(\tau,\s)$ is itself periodic. Varying the total metastring action we get

\begin{small}
\be \delta S_{T} = \frac1{2\pi} \int_{\tau_{i}}^{\tau_{o}} \rd\tau \int_{\s_{0}}^{\s_{2\pi}} \rd\s \pa_{\s} \S + \frac1{4\pi} \Big [ \int_{\s_{0}}^{\s_{2\pi}} \rd\s \delta \X (\eta+\omega) \pa_{\s}\X \Big |_{\tau_{i}}^{\tau_{o}} + \int_{\tau_{i}}^{\tau_{o}} \rd\tau \delta \X  [2\eta \S - (\eta+\omega) \pa_{\tau} \X] \Big |_{\s_{0}}^{\s_{2\pi}} \Big ]
\ee

\end{small}

\begin{figure}
	\centering
	\includegraphics[scale=0.5]{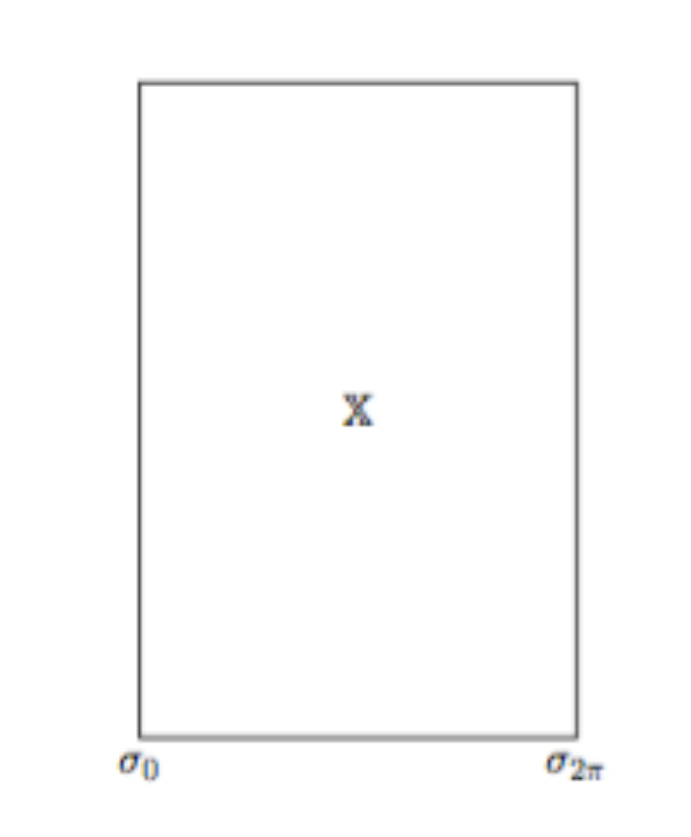}\label{fig1}
	\caption{Cutting open a cylindrical world-sheet: One-strip decomposition}
	\end{figure}

where

\be \S \equiv \pa_{\tau}\X - J\pa_{\s}\X \ee

The first term in equation (2.20) is the bulk term from which we extract the bulk equations of motion:

\be \pa_{\s}\S = \pa_{\s}(\pa_{\tau} - J \pa_{\s} ) \X = 0. \ee

As in [2], we introduce the center of mass coordinate $x$ and the monodromy $\Delta$:

\be \oint \rd\s \X(\tau, \s) = x(\tau), \hspace{10pt }\oint \rd\s \pa_{\s} \X(\tau, \s) = \frac{\Delta(\tau)}{2\pi} , \ee

where $\oint \equiv \frac1{2\pi} \int_{0}^{2\pi}.$ By using the equations of motion and the periodicity of $\pa_{\s} \X $, we get the following expressions and notice that $\Delta$ is time independent.

\be \label{S} \S(\tau) = \oint \S(\tau) = \pa_{\tau}x - \frac{J\Delta}{2\pi}. \ee
\be 0 = \oint \pa_{\s} \S = \frac{\pa_{\tau}\Delta}{2\pi}. \ee

For convenience, let us introduce the coordinates

\be \bar{x}(\tau) \equiv \int_{0}^{\tau} d\tau \S(\tau). \ee

Using the above, the time dependence of the centre of mass position can be written as 

\be  x(\tau) = x_{c} + \frac{J\Delta}{2\pi}\tau + \bar{x}(\tau). \ee

where, $x_{c}$ is time-independent. Using equation (\ref{S}), we can write the general solution to the equation of motion

\be \X(\tau, \s) = x_{c} + \frac{J\Delta}{2\pi}\tau +  \frac{\Delta}{2\pi}\s + \Q(\tau, \s) + \bar{x}(\tau), \ee

where $\Q$ is chiral: $(\pa_{\tau} - J\pa_{\s}) \Q(\tau, \s) = 0$ and periodic under $\s \rightarrow \s + 2\pi$. Consequently,  it is possible to fourier expand

\beqn
{\Q}(\tau,\sigma)&=&\sum_{n\in\mathbb{Z}_*}\Q^{+}_n e^{-in(\tau+\sigma)}
-\sum_{n\in\mathbb{Z}_*}\Q^{-}_n e^{-in(\tau-\sigma)}\\
&\equiv & {\Q}^{+}(\tau+\sigma)- {\Q}^{-}(\tau-\sigma).
\eeqn

where ${\Q}^{\pm}$  are identified with equations (\ref{Jleft}) and (\ref{Jright}) as the left- and right- movers in the usual string:

 \be
 J{\Q}^{+}=+{\Q}^{+},\qquad J{\Q}^{-}=-\Q^{-}.
 \ee

\section{Symplectic Structure: a Discussion}

Let us begin our discussion with the question: what does it mean to glue two strip-edges together? In the symplectic language, it is analogous to saying that the symplectic flux across the boundary is conserved. Let us review the extraction of the symplectic potential as done in [2] as a model to follow for the $N$ -strip calculation we will be attempting in chapter 4. The total action can be written in the form

\be\delta S_{T} \hat{=} \Theta(\tau_{o}) - \Theta(\tau_{i}),  \ee

where $S_{T}$ is the total action  for any number of strips. In order to extract the form of the symplectic potential and the symplectic form from the variation of the action, we introduce the notion of a differential on field space denoted as  $\delta$ which satisfies the Leibniz rule and squares to zero, i.e.  $\delta^2 = 0$. Reviewing section (4.2) in [2], we see that in order to conserve the symplectic flux across the cuts, we must introduce some boundary term variations in the action. This promotes some boundary data associated with the cuts to dynamical degrees and freedom and compensates for the symplectic flux across the cuts. Also, since the theory is conformal, there should ideally be no dependence of the physics on the position of each cut as well the orientation of the strips. The one strip action, equation (2.19), along with the proposed boundary term (keeping the indices implicit) is

\beq\label{actionpatch}
S_T=\frac1{4\pi} \int_{\tau_o}^{\tau_i}\rd \tau \int_{\s_{0}}^{\s_{0}+2\pi} \rd \s\left[  \pa_{\tau}\X (\eta+ \omega)\pa_{\s}\X -  
  \pa_{\s}\X H \pa_{\s}\X\right] - s(\s_{0}),
\eeq

where  $\s_{0}$ is the position of the cut. Demanding that the action be independent of the position of  $\s_{0}$ and remains invariant under the change of orientation $\s \rightarrow 2\pi - \s $, we find, 

\be s(\s_{0}) = \frac{1}{8\pi} \int_{\tau_{i}}^{\tau_{o}} \rd \tau \pa_{\tau} \Delta (\eta + \omega) (\X_{\s_{0}} + \X_{\s_{0}+ 2\pi} ) \ee

satisfies our demands. Now, since we have made our theory conformally invariant and independent of the position of the cut, we can fix it anywhere we like. For simplicity, we stick to the cut position to be (0, $2\pi$) and looking at equation (2.20), we see that the total variation of the action can be written as

\begin{align} \delta S_{T} = \frac1{2} \Big [ \oint \rd \s \delta \X(\eta+\omega)\pa_{\s}\X \Big ]_{\tau_{i}}^{\tau_{o}} &- \frac1{8\pi} [\delta \Delta (\eta+\omega) (\X_{0} + \X_{2\pi}) ]_{\tau_{i}}^{\tau_{o}}  + \frac1{2\pi} \int_{\tau_{i}}^{\tau_{o}} \rd \tau \delta \X\eta\S \Big |_{0}^{2\pi} \nonumber \\
\qquad &- \frac1{4\pi} \int_{\tau_{o}}^{\tau_{i}} \rd \tau \Big( \delta (\X_{0} + \X_{2\pi} ) \eta \pa_{\tau} \Delta + 2 \int \delta \X\eta\pa_{\s} \S \Big ) . \end{align}

The last term, which is the bulk variation term of the action tells us that we are still off-shell. Remembering that we can extract the symplectic form from the on-shell variation of the action, the equations of motion (2.25) and (2.22) are imposed and using equations (2.32) and (2.26), we extract the symplectic form to be

\be \Theta = \frac1{2} \oint \rd \s (\delta \X (\eta + \om) \pa_{\s} \X) - \frac1{8\pi} \delta \Delta (\eta+\om) (\X_{0} + \X_{2\pi}) + \frac1{2\pi} \delta \Delta \eta \bar{x}. \ee

\section{Symplectic Potential}

The symplectic potential can be calculated from equation (2.36), by taking the variation $\Omega =  \delta \Theta $ and as in [2], we get

\be \Omega = \frac1{2} \oint \rd \s \delta \X \cdot \pa_{\s} \delta\X - \frac1{4\pi} \delta \Delta \cdot \delta\X_{0} + \frac1{2\pi} \delta \Delta \cdot \delta \bar{x}. \ee

The above expression looks very simple. Let us first take note that any reference to the end points of the strip or the position of the cut has vanished. Now, this is conformally invariant and only depends on the cylinder topology as it should. Moreover, we also notice that our expression is also independent of the the two-form $\omega$ which means that despite the cut, the string is closed since the term proportional to $\omega$ is a closed form.

\begin{savequote}[45mm]
Everything we call real is made of things that cannot be called as real.
\qauthor{Niels Bohr}
\end{savequote}

\chapter{Two Strip Analysis}

\begin{figure}[ht]
	\centering
	\includegraphics[scale=0.5]{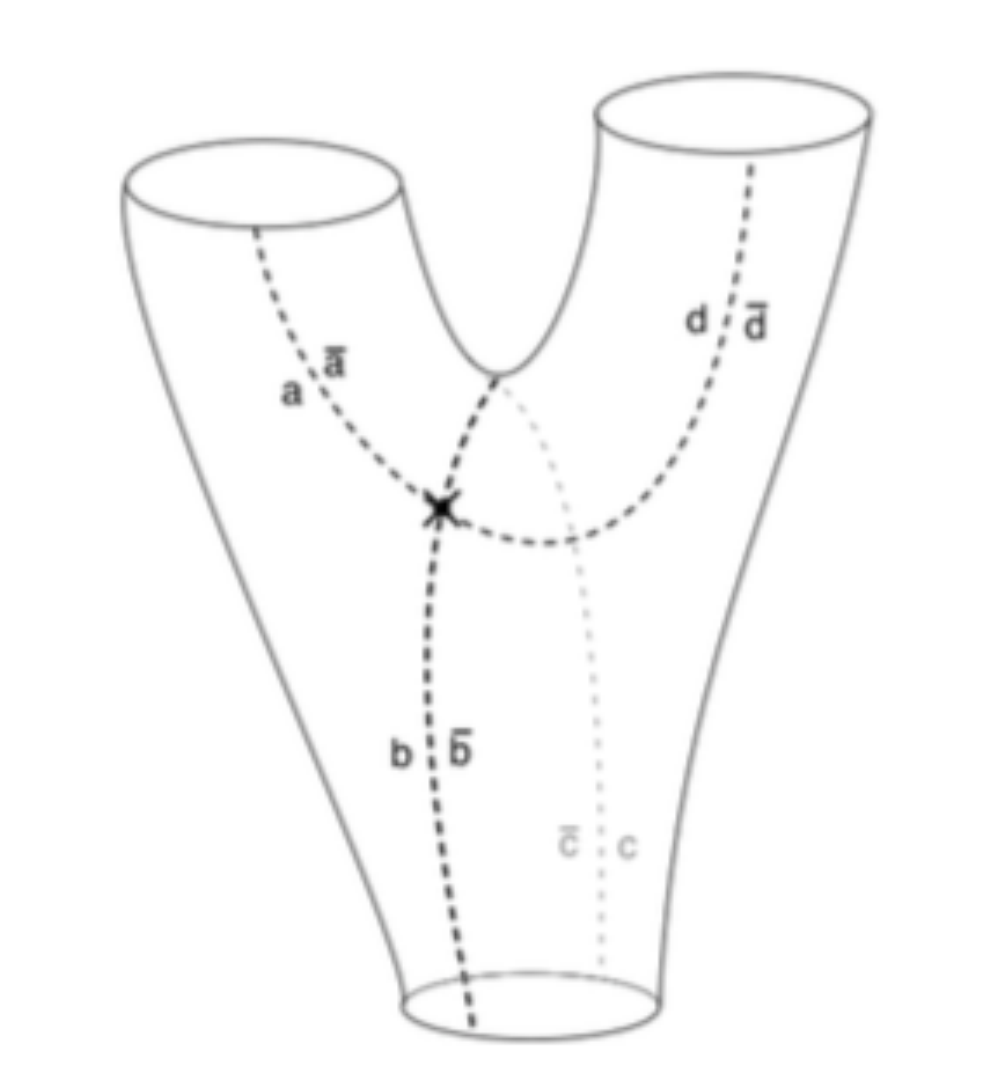}\label{fig2}
	\caption{Pants Diagram in String Theory [2]}
	\end{figure}

Looking at Figure 2.1, we see that a tree level simple closed string propagation cylinder can be decomposed into a single strip with a single cut. Thus, it was sufficient to study the gluing condition for a single strip. But now, let us consider Figure 3.1. The string pants diagram has one interaction point. It cannot be decomposed into a strip picture with a single cut, we require a minimum of two strips to stitch the pants as shown in Figure 3.2. Thus, we can form any kind of string interaction picture such as Figure 4.1(a) as a decomposition and gluing of many such strips.
In order to be able to analyse multi-nakamura strips, we must go step by step. In this section we focus on the gluing conditions of two strips and try to determine the equations of motion of the boundaries. Later, we will analyse the symplectic structure and the flux and generalize our analysis to the gluing of general Nakamura strips. We follow the analysis of gluing one strip back together as shown in [2]. But for ease of calculation, we introduce a new notation: We fix the positive direction as shown by the arrows in Figure 3.2. Let us name the strips 1 and 2 and denote the respective coordinates on them by $\X^1$ and $\X^2$. We define the monodromies of the two strips as seen from Figure 3.2.

\be \Delta_{12} = \X^{1}_{12} - \X^{2}_{12}, \ee

\begin{figure}
	\centering
	\includegraphics[scale=0.5]{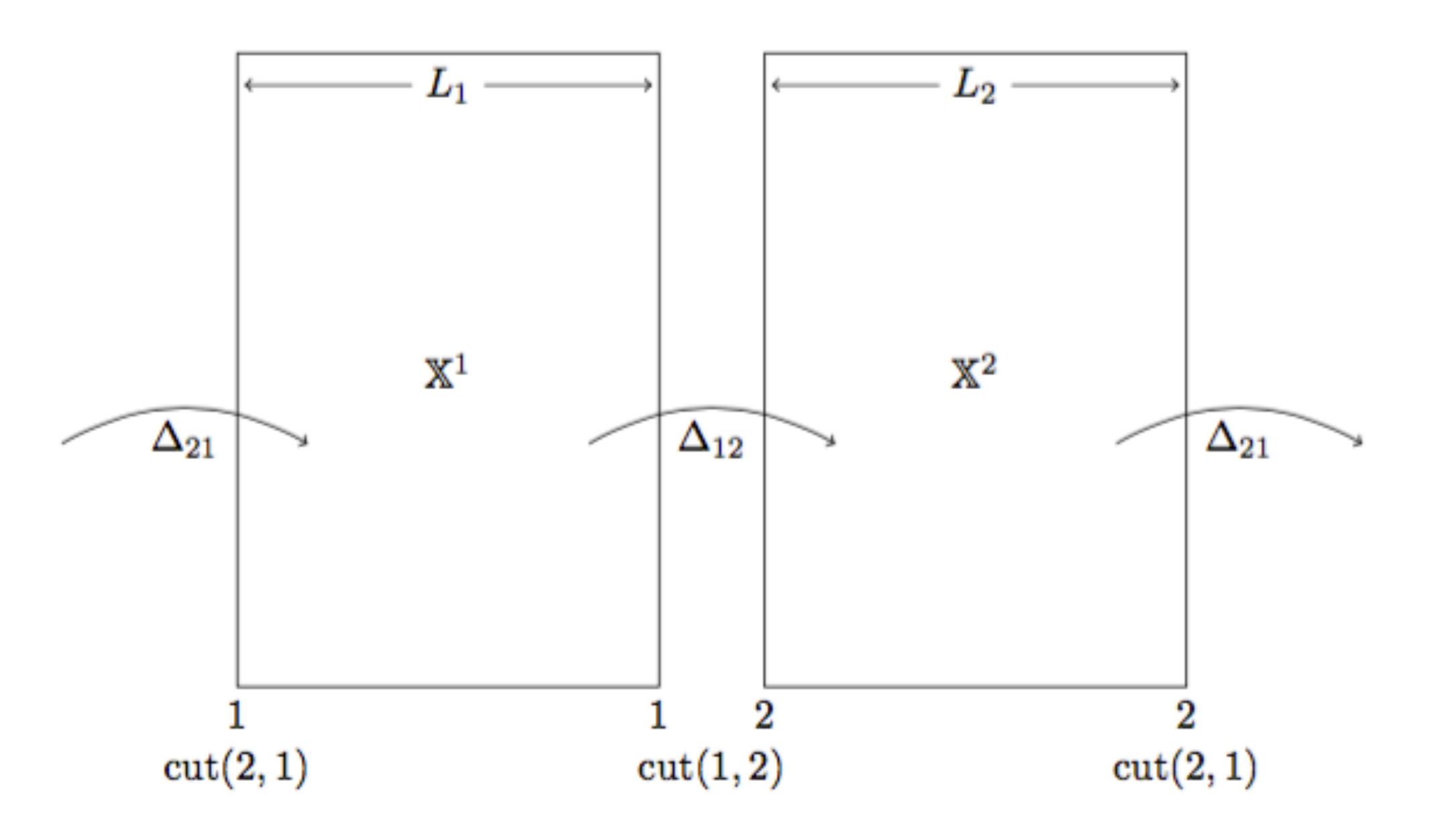}\label{fig3}
	\caption{Two strip decomposition}
	\end{figure}

which is the monodromy across cut(1,2) and

\be \Delta_{21} = \X^{2}_{21} - \X^{1}_{21}, \ee

is the monodromy across cut (2,1). Adding the two equations (3.1) and (3.2) we get

\be \Delta = \Delta_{12} + \Delta_{21} = \X^{1}_{12} - \X^{2}_{12} + \X^{2}_{21} - \X^{1}_{21}, \ee

which is the total monodromy and should be conserved. Looking at Figure 3.2, we can also write the two strip length or strip separation vectors as,

\be L_{1} = \X^{1}_{12} - \X^{1}_{21} \ee

and 

\be L_{2} = \X^{2}_{21} - \X^{2}_{12}. \ee

Similarly adding the two lengths, we get

\be L = L_{1} + L_{2} = \X^{1}_{12} - \X^{1}_{21} + \X^{2}_{21} - \X^{2}_{12} = \Delta, \ee

and L is the total length. The multistrip action can be written as the one-strip action summed over the
number of strips, in this case, $N = 2$.

\be S_{T} = \frac1{4\pi} \sum_{i} \int_{S_{i}} \rd \tau \rd \s \big [ \pa_{\tau} \X^{A}_{i} (\eta_{AB} + \omega _{AB}) \pa_{\s}\X_{i}^{B} - \pa_{\s} \X_{i}^{A} H_{AB} \pa_{\s}\X_{i}^{B} \Big ] \ee

\section{The Bulk Equations of Motions}

We recover the equations of motion from the bulk variation of the action (see equation 2.22) as follows:

\be \sum_{i}  \pa_{\s} \S^{i} = 0 \ee
 where 
 \be 
\S^{i} = \pa_{\tau}\X^{i} - J\pa_{\s}\X^{i}. \ee

The continuity across the cuts implies that across the cuts (1,2) and (2,1) respectively, we impose

\be \pa_{\s} \X^{1}_{12} = \pa_{\s} \X^{2}_{12} , \qquad \pa_{\s} \X_{21}^{1} = \pa_{\s} \X_{21}^{2}. \ee

Let us now introduce, 
\be l_{1} = \s_{12} - \s_{21} \ee

which is the strip length of the first strip. Note that $l_1$, which is the strip length is different from $L_1$ which is a vector and depends on both $\s$ and $\tau$. But once we have this distinction in mind, we will just use strip length to refer to $L_1$ in this essay. Moreover, by integrating the equation of motion for the first strip, and using equation (3.11), we see that

\be \frac1{l_{1}} \int_{21}^{12} \S^{1} = \S^{1}_{12} - \S^{1}_{21} = \frac1{l_{1}} (\pa_{\tau} L_{1} - J \pa_{\s}L_{1} ) = 0. \ee

The reader might be confused by the notation $\pa_{\s}L_1$. If we look at equation (3.4) and write $L_1 \equiv \X^1(\s_{21}) - \X^1(\s_{12})$ which is a function of $\s_{12}$ and $\s_{21}$. We can now translate both $\s_{12}$ and $\s_{21}$ by $\s$ such that we now compute and denote, $\pa_{\s}L_1 \equiv \pa_{\s}L_1(\s_{12}+\s , \s_{21}+\s)$ which makes it clear. The first part of equation 3.12 tells us that

\be \S^{1}_{12} = \S^{1}_{21} = \S^{1}. \ee

Using the above, we can compute

\begin{align}
\pa_{\tau}\Delta_{12} &= \pa_{\tau}[\X^{1}_{12} - \X^{2}_{12} ], \nonumber \\
\qquad & = \pa_{\tau}\X^{1}_{12} - \pa_{\tau}\X^{2}_{12}, \nonumber \\
\qquad & = \S^{1}_{12} + J\pa_{\s}\X^{1}_{12} - \S^{2} _{12} - J\pa_{\s}\X^{2}_{12}, \nonumber \\
\qquad & = \S^{1} - \S^{2}.
\end{align} 

Similarly, we can compute

\be \pa_{\tau}\Delta_{21} = \S^{2}-\S^{1}, \ee

which gives us a relationship between the strip monodromies, i.e, 

\be \pa_{\tau} \Delta_{12} = -\pa_{\tau} \Delta_{21} \ee

and subsequently a relationship between the strip length vectors, 

\be \Rightarrow \pa_{\tau}L_{1} = -\pa_{\tau}L_{2}. \ee

Now, we denote

\be \frac1{l_{1}}\int_{21}^{12} \rd \s \X^{1} = x_{1}(\tau), \ee

as the centre of mass coordinate of the strip one. Similarly for strip two we have

\be \frac1{l_{2}}\int_{12}^{21} \rd \s \X^{2} = x_{2}(\tau), \ee

where $l_{2} = \s_{21} - \s_{12}$. Furthermore, we also have

\be  \frac1{l_{1}}\int \rd \s \pa_{\s} \X^{1} = L_{1}. \ee

Using the above, we calculate

\begin{align}
\S^{1}(\tau) \equiv \frac1{l_{1}} \int_{21}^{12} \rd \s \S^{1} (\s, \tau) &=  \frac1{l_{1}} \int_{21}^{12} \rd \s (\pa_{\tau} - J \pa_{\s} ) \X^{1}, \nonumber \\
\qquad &= \pa_{\tau} x_{1} - \frac1{l_{1}} J \X^{1} \Big |_{21}^{12} \nonumber \\
\qquad &= \pa_{\tau} x_{1} - \frac1{l_{1}} J L_{1}.
\end{align}

It would be easier and neater to have our equations depend on a fixed set of the same basis variables. In order to come up with a basis set of variables, we denote $\X^{1}_{12} = \X^{1}$ and $\X^{2}_{21} = \X^{2}$. We use $(\X^{1}, \X^{2}, \Delta, \Delta_{12} \hspace{2pt}\text{and} \hspace{2pt}\S^{2})$ as our basis variables and express all other variables in terms of these as follows:

\begin{align} \X^{1}_{21} &= \X^{2}_{21} - \Delta_{21} = \X^{2} + \Delta_{12} - \Delta, \nonumber \\ 
\qquad \X^{2}_{12} &= \X^{1}_{12} - \Delta_{12} = \X^{1} - \Delta_{12} \nonumber \\  \text{and}  \hspace{5pt} \qquad  \S^{1} &= \pa_{\tau}\Delta_{12} + \S^{2} \end{align}

\section{Analysis of the boundaries}

Following Figure 3.2, we have named the two strips 1 and 2, with generalized coordinates $\X^{1}$ and $\X^2$ respectively. $\X^{1}(1,2)$ at the cut (1,2) is denoted as $\X^{1}_{12}$ and so on. Consider the boundary terms in the variation of the two strip action. As before, looking at equation (2.20), we see that the on-shell variation of the action is a pure boundary term in terms of the two cuts (1,2) and (2,1)

\be \delta S \hat{=}   \frac1{4\pi} \int_{\tau_{i}}^{\tau_{o}} \rd \tau \Big [  \delta \X^{1}  (2 \eta \S^{1} - (\eta + \omega) \pa_{\tau} \X^{1}) \Big ]\Big |_{\s_{21}}^{\s_{12}} + \frac1{4\pi}  \int_{\tau_{i}}^{\tau_{o}} \rd \tau \Big [ \delta \X^{2}  (2\eta \S^{2} - (\eta+\omega) \pa_{\tau} \X^{2}) \Big] \Big |_{\s_{12}}^{\s_{21}}.
\ee

Here, we have only considered the $\tau$ integral for the time being, to keep things simple and $\hat{=}$ denotes that the action is on-shell. We can now proceed to write the variation in terms of the monodromies and $\pa_{\tau}\X$'s

\begin{align} \delta S &\hat{=}   \frac1{4\pi} \int_{\tau_{i}}^{\tau_{o}} \rd \tau \Big [\Big[ \delta \X^{1}_{12}  (2 \eta \S^{1} - (\eta + \omega) \pa_{\tau} \X^{1}_{12}) \Big ] - \Big [ \delta \X^{1}_{21}  (2 \eta \S^{1} - (\eta + \omega) \pa_{\tau} \X^{1}_{21}) \Big ]  +  \nonumber \\ 
\qquad & \Big [ \delta \X^{2}_{21}  (2 \eta \S^{2} - (\eta + \omega) \pa_{\tau} \X^{2}_{21}) \Big ] - \Big [ \delta \X^{2}_{12}  (2 \eta \S^{2} - (\eta + \omega) \pa_{\tau} \X^{2}_{12}) \Big ]\Big]. \end{align}

Examining the expression above, we see terms depending on the cut (2,1) and (1,2) respectively. Collecting the terms depending on the cut (1,2) we get the action at cut (1,2) denoted by $S_{12}$ as

\be \delta S_{12} \hat{=}   \frac1{4\pi} \int_{\tau_{i}}^{\tau_{o}} \rd \tau \Big [\Big[ \delta \X^{1}_{12}  (2 \eta \S^{1} - (\eta + \omega) \pa_{\tau} \X^{1}_{12}) \Big ] - \Big [ \delta \X^{2}_{12}  (2 \eta \S^{2} - (\eta + \omega) \pa_{\tau} \X^{2}_{12}) \Big ]\Big]. \ee

Using equations (3.1), (3.4) and (3.14)

\begin{align} \delta S_{12} &\hat{=}   \frac1{4\pi} \int_{\tau_{i}}^{\tau_{o}} \rd \tau\Big[ \Big[ \delta \X^{1}_{12}  (2 \eta \S^{1} - (\eta + \omega) \pa_{\tau} \X^{1}_{12}) \Big ] - \Big [ ( \delta \X^{1}_{12} - \delta \Delta_{12}) (2 \eta \S^{2} - (\eta + \omega) \pa_{\tau} \X^{2}_{12}) \Big ] \Big ], \nonumber \\
\qquad &\hat{=}  \frac1{4\pi} \int_{\tau_{i}}^{\tau_{o}} \rd \tau \Big[ \delta \X^{1}_{12}  (2 \eta \S^{1}  - 2 \eta \S^{2} - (\eta + \omega) \pa_{\tau}\X^{1}_{12} + (\eta + \omega) \pa_{\tau}\X^{2}_{12} ) + \delta \Delta_{12} (2\eta \S^{2} - (\eta + \omega) \pa_{\tau}\X^{2}_{12} ) \Big ], \nonumber \\
\qquad &\hat{=}  \frac1{4\pi} \int_{\tau_{i}}^{\tau_{o}} \rd \tau \Big[ \delta \X^{1}_{12} (2\eta \pa_{\tau}\Delta_{12} - (\eta + \omega) \pa_{\tau}\Delta_{12}) + \delta \Delta_{12} (2\eta \S^{2} - (\eta + \omega) \pa_{\tau}\X^{2}_{12}) \Big ], \nonumber \\
\qquad &\hat{=}  \frac1{4\pi} \int_{\tau_{i}}^{\tau_{o}} \rd \tau \Big[ \delta \X^{1}_{12}(\eta - \omega) \pa_{\tau}\Delta_{12}) + \delta \Delta_{12} (2\eta\S^{2} - (\eta + \omega) \pa_{\tau}\X^{2}_{12}) \Big]. 
\end{align}

Similarly, the variation of the action for the cut (2,1) can be written as

\be \delta S_{21} \hat{=} \frac1{4\pi} \int_{\tau_{i}}^{\tau_{o}} \rd \tau \Big[ \delta \X^{2}_{21}(\eta - \omega) \pa_{\tau}\Delta_{21} + \delta \Delta_{21} (2\eta\S^{1} - (\eta + \omega) \pa_{\tau}\X^{1}_{21}) \Big]. \ee

Going to the $(\X^{1}, \X^{2}, \Delta_{12}, \Delta \hspace{2pt }\text{and} \hspace{2pt} \S^2)$ basis determined above

\begin{align} \delta S_{12} &\hat{=}  \frac1{4\pi} \int_{\tau_{i}}^{\tau_{o}} \rd \tau \Big[ \delta \X^{1}_{12}(\eta - \omega) \pa_{\tau}\Delta_{12} + \delta \Delta_{12} (2\eta\S^{2}_{12} - (\eta + \omega) \pa_{\tau}\X^{2}_{12}) \Big]. \nonumber \\
\qquad &\hat{=}  \frac1{4\pi} \int_{\tau_{i}}^{\tau_{o}} \rd \tau \Big[ \delta \X^{1}(\eta - \omega) \pa_{\tau}\Delta_{12} + \delta \Delta_{12} (2\eta\S^{2} - (\eta + \omega)( \pa_{\tau}\X^{1} - \pa_{\tau}\Delta_{12})) \Big], \nonumber \\
\qquad &\hat{=}  \frac1{4\pi} \int_{\tau_{i}}^{\tau_{o}} \rd \tau \Big[ \delta \X^{1}(\eta - \omega) \pa_{\tau}\Delta_{12} + \delta \Delta_{12} (2\eta\S^{2} - (\eta + \omega)\pa_{\tau}\X^{1} +  (\eta + \omega) \pa_{\tau} \Delta_{12} ) \Big]. \end{align}

and similarly,

\begin{align} 
\delta S_{21} &\hat{=} \frac1{4\pi} \int_{\tau_{i}}^{\tau_{o}} \rd \tau \Big[ \delta \X^{2}_{21}(\eta - \omega) \pa_{\tau}\Delta_{21}) + \delta \Delta_{21} (2\eta\S^{1}_{21} - (\eta + \omega) \pa_{\tau}\X^{1}_{21}) \Big]. \nonumber \\
\qquad &\hat{=} \frac1{4\pi} \int_{\tau_{i}}^{\tau_{o}} \rd \tau \Big [ - \delta \X^{2}(\eta - \omega) \pa_{\tau}\Delta_{12} + (\delta\Delta - \delta \Delta_{12}) (2 \eta \S^{1} - (\eta + \om) \pa_{\tau} \X^{1}_{21} )\Big ], \nonumber \\
\qquad &\hat{=} \frac1{4\pi} \int_{\tau_{i}}^{\tau_{o}} \rd \tau \Big [ - \delta \X^{2}(\eta - \omega) \pa_{\tau}\Delta_{12} + (\delta\Delta - \delta \Delta_{12}) (2 \eta (\pa_{\tau}\Delta_{12}+\S^{2}) - (\eta+\omega) \pa_{\tau}\X^{1}_{21}) \Big ], \nonumber \\
\qquad &\hat{=} \frac1{4\pi} \int_{\tau_{i}}^{\tau_{o}} \rd \tau \Big [ - \delta \X^{2}(\eta - \omega) \pa_{\tau}\Delta_{12} + \delta \Delta (( \eta - \omega ) \pa_{\tau} \Delta_{12} + 2\eta \S^2 -  (\eta + \omega) \pa_{\tau} \X^2)  \nonumber \\
\qquad & \hspace{20pt} + \delta \Delta_{12} (-(\eta-\omega) \pa_{\tau} \Delta_{12} - 2 \eta \S^2 + (\eta + \omega ) \pa_{\tau}\X^2 ) \Big ]. 
\end{align}

where we have made use of the fact that $\pa_{\tau}\Delta = 0$. Recombining $\delta S_{12}$ and $\delta S_{21}$, the total variation looks like

\begin{align} \delta S \hat{=} \frac1{4\pi} \int_{\tau_{i}}^{\tau_{o}} \rd \tau \Big [ (\delta \X^1- \delta \X^2) (\eta - \omega) \pa_{\tau}\Delta_{12} &+ \delta \Delta (( \eta - \omega) \pa_{\tau}\Delta_{12} + 2\eta\S^{2} - (\eta+\omega) \pa_{\tau}\X^2) \nonumber \\ 
\qquad &+  \delta\Delta_{12}(2\omega \pa_{\tau}\Delta_{12} + (\eta+\omega)(\pa_{\tau}\X^2 - \pa_{\tau}\X^1)) \Big ]. 
\end{align}

Demanding that this variation vanishes at the boundaries, leads us to the following boundary equations of motion:

\be \delta \X : (\eta- \om) \pa_{\tau} \Delta_{12} = (\eta- \om) \pa_{\tau} \Delta_{21} = 0, \ee
\be \delta \Delta : 2 \eta \S^{2} = (\eta + \om) \pa_{\tau} \X^{2}, \ee

and 

\be \delta \Delta_{12} : (\eta + \om) (\pa_{\tau} \X^{2} - \pa_{\tau}\X^{1}) + 2\om\pa_{\tau}\Delta_{12} = 0. \ee

Massaging the third equation to a form we can interprete in terms of the evolution of the boundary,

\be (\eta + \om) \pa_{\tau} L_{1} = (\eta - \om) \pa_{\tau} \Delta_{21} = 0 \ee
\be (\eta - \om) \pa_{\tau} \Delta_{12} = (\eta + \om) \pa_{\tau} L_{2} = 0 \ee

Thus, finally we have extracted the following equations of motion for the boundary:

\be (\eta + \om) \pa_{\tau} L_{2} = 0 \Rightarrow (\eta+\om)\pa_{\tau} L = 0 \ee
\be (\eta - \om) \pa_{\tau} \Delta_{12} = 0 \Rightarrow  (\eta - \om) \pa_{\tau} \Delta = 0 \ee

\be 2\eta \S^{2} =  (\eta + \om) \pa_{\tau} \X^{2}_{21} \ee
\be 2\eta \S^{1} =  (\eta + \om) \pa_{\tau} \X^{1}_{21} \ee

Moreover, we already had the following bulk equations of motion,

\be \pa_{\tau}L = 0.  \ee
\be \pa_{\tau} \Delta = 0. \ee

\section{Stripology, String-Bits and Chan-Paton Factors}

Now, at this juncture, let us pause a while and discuss the potential consequences of this strip picture. We can see from the equations (3.36) and (3.37) that we can easily generalize this 2-strip picture to $N$ -strips. The study of the strip-picture or \textit{``stripology''}, as we call it, is important in the sense that it paves a way to naturally combine the Chan-Paton factors that come from matrix models of string theory as well as the string bits picture proposed by Klebanov and Susskind in [3] and [4]. In our case, instead of string bits, we will deal with strip bits as the building blocks. Moreover, we conjecture that we can associate a symplectic variable $\alpha$ at each cut, and hope that we can manage to write the strip amplitude as a product of strip matrices depending on these symplectic variables $\alpha$, $\beta$ etc.

\section{Open Strings, and Meta-D-Branes: a Discussion}

In the above treatment, we varied the boundary action of the metastrips and demanded that the variation vanishes. However, that is not what happens when we want to glue the strings back together. Let us again summon the equations of motion, (3.36) and (3.37) and try to make sense of them.

\be (\eta+\om) \pa_{\tau} L_{2} = 0 \ee
\be  (\eta-\om) \pa_{\tau} \Delta_{12} = 0 \ee

Demanding that the variation vanishes at the boundary effectively truncates the boundary. Looking at the above equations, we recognize that these are reminiscent of the Dirichlet boundary conditions at the two ends of the strip. Thus, we have open strips (open-string propagation) ending on meta-D-branes. In the present formalism, we see a very natural evolution of the D-branes and call these the meta-D-branes. However, the D-branes prevent gluing the strips back together. Hence, if we want to glue the strips back together and form a closed strip worldsheet, we must treat the variation differently. We conjecture that there must be some kind of momentum conservation equation which makes sure the symplectic flux is conserved at the cut. Note, however, that vanishing of the symplectic flux at the boundary does not imply that the strips are glued together. It might also happen in case of open strips where there is no flux flow out through the boundaries. The next chapter deals with closed strips and the symplectic structure rigorously. We will also proceed to generalize the above procedure to $N$ -strips.

\begin{savequote}[45mm]
Everything should be as simple as possible, but not simpler.
\qauthor{Albert Einstein}
\end{savequote}

\chapter{Boundary Symplectic Structure}

The previous chapter saw the natural evolution of meta-D-branes associated with the open strip boundary conditions from the metastring theory formalism. In this chapter, we will proceed to rigorously examine the gluing condition for the 2-cuts in the 2 strip picture in Figure 3.2. Following the discussion in section (2.5), we proceed to add a boundary term to each cut (1,2) and (2,1) in the two-strip action such that all the symplectic flux through the cuts are compensated. This will show that the cuts can be glued back together and the string is closed. From, the general meta-string action in equation (3.7), we see that the two-strip action can be written explicitly as

\begin{align}
S_{T} &= \frac1{4\pi} \int_{\tau_{i}}^{\tau_{o}} \rd \tau  \int_{\s_{21}}^{\s_{12}} \rd \s 
\Big [ \pa_{\tau} \X^{1} (\eta + \om) \pa_{\s} \X^{1} - \pa_{\s} \X^{1} H \pa_{\s} \X^{1} \Big ] - s_{12} \nonumber \\
\qquad &+  \frac1{4\pi} \int_{\tau_{i}}^{\tau_{o}} \rd \tau  \int_{\s_{12}}^{\s_{21}} \rd \s 
\Big [ \pa_{\tau}\X^{2} (\eta + \om) \pa_{\s} \X^{2} - \pa_{\s} \X^{2} H \pa_{\s} \X^{2} \Big ] - s_{21},
\end{align}

where $s_{12}$ and $s_{21}$ respectively, are the extra boundary terms added at the cuts (1,2) and (2,1) respectively. To preserve conformal invariance, we demand that the action be independent of the position of the cuts, i.e independent of $\s_{12}$ and $\s_{21}$ and invariant under the change of orientation. At the cut (1,2), we must demand

\be \pa_{\s_{12}} S_{T} = 0, \ee

which implies,

\begin{small}
\be  \frac1{4\pi} \int_{\tau_{i}}^{\tau_{o}} \rd \tau \Big [ \pa_{\tau} \X^{1}_{12} (\eta + \om) \pa_{\s_{12}} \X^{1}_{12} - \pa_{\s_{12}} \X^{1}_{12} H \pa_{\s_{12}} \X^{1}_{12} - 
\pa_{\tau} \X^{2}_{12} (\eta + \om) \pa_{\s_{12}} \X^{2}_{12} + \pa_{\s_{12}} \X^{2}_{12} H \pa_{\s_{12}} \X^{2}_{12} \Big ] = \pa_{\s_{12}}s_{12}. \ee

\end{small}

Using the continuity equations (3.10), we are left with

\begin{align}  \frac1{8\pi} \int_{\tau_{i}}^{\tau_{o}} \rd \tau \Big [ \pa_{\tau} (\X^{1}_{12}-\X^{2}_{12})  (\eta + \om) \pa_{\s_{12}} (\X^{1}_{12}+ \X^{2}_{12}) &= \pa_{\s_{12}} s_{12}, \nonumber \\
\Rightarrow   \frac1{8\pi} \int_{\tau_{i}}^{\tau_{o}} \rd \tau \Big [ \pa_{\tau} \Delta_{12}
 (\eta + \om) \pa_{\s_{12}} (\X^{1}_{12}+ \X^{2}_{12}) &= \pa_{\s_{12}} s_{12}. \end{align}
 
We know that $\Delta_{12}$ is independent of $\s_{12}$, hence we can integrate out $\s_{12}$ and recover the boundary terms that must be added for the cut (1,2) as follows:

\be  s_{12} =  \frac1{8\pi} \int_{\tau_{i}}^{\tau_{o}} \rd \tau \Big [ \pa_{\tau} \Delta_{12}
 (\eta + \om) (\X^{1}_{12}+ \X^{2}_{12}) \Big ]. \ee

Similarly, at the cut (2,1), we must add a term

\be  s_{21} =  \frac1{8\pi} \int_{\tau_{i}}^{\tau_{o}} \rd \tau \Big [ \pa_{\tau} \Delta_{21}
 (\eta + \om) (\X^{1}_{21}+ \X^{2}_{21}) \Big ]. \ee
 
Looking at equations, (4.5) and (4.6), let us collect the terms for each strip. For example, the first strip $S^1$ will have terms added to each boundary (1,2) and (2,1). We collect them into $s^1$.

\begin{align}
s^1& = s^1_{12} + s^1_{21} \nonumber \\
\qquad &=   \frac1{8\pi} \int_{\tau_{i}}^{\tau_{o}} \rd \tau \Big [ \pa_{\tau} \Delta_{12}
 (\eta + \om) \X^{1}_{12} +  \pa_{\tau} \Delta_{21}
 (\eta + \om) \X^{1}_{21} \Big ], \nonumber \\
\qquad &=   \frac1{8\pi} \int_{\tau_{i}}^{\tau_{o}} \rd \tau \pa_{\tau} \Delta_{12}
 (\eta + \om) (\X^{1}_{12}  - \X^{1}_{21}), \nonumber \\
\qquad &=   \frac1{8\pi} \int_{\tau_{i}}^{\tau_{o}} \rd \tau  \pa_{\tau} \Delta_{12}
 (\eta + \om) L_{1}, 
\end{align}

 where we have used equations (3.4) and (3.16) off-shell. Similarly the terms necessary for strip 2 are
 
 \be s^{2} = s^2_{12} + s^2_{21} =  \frac1{8\pi} \int_{\tau_{i}}^{\tau_{o}} \rd \tau  \pa_{\tau} \Delta_{21}
 (\eta + \om) L_{2}. \ee
 
 At this point, we have successfully removed the dependence of the action on the position of the cuts. For simplicity, we can now safely let all the strips have the same width and invoking conformal invariance, we can rescale each of the strips to run from $[0, 2\pi]$. Thus, we can set $l_2 = l_1 = 2\pi$. Thus, now each strip in the two strip picture is a copy of the one-strip in the one-strip picture (Figure 2.1), which makes it possible to use the same expressions.

\section{The Symplectic Potential}

The symplectic potential can be extracted from the on-shell variation of the total action as given in equation (2.32) as

\be \delta S_{T} \hat{=} \Theta(\tau_{o}) - \Theta(\tau_{i}). \ee

Thus, we proceed to calculate the variation of the total action. Looking at equation (4.1), we can write
the variation as a variation over each strip separately as

\be \delta S_{T} =  \delta S_{T}^1 +  \delta S_{T}^2 =  \delta S^1 - \delta s^1 +  \delta S^2 - \delta s^2. \ee

Separating the terms for Strip 1:

\be  \delta S_{T}^1 = \delta S^1 - \delta s^1 \ee

where, $S_{T}^{1}$ is the total variation of the first strip-action including the additional boundary term, $\delta s^1$ are the extra-boundary variation terms for both the boundaries (1,2) and (2,1) and $S^1$ is the full action for the first strip (see equation 2.20) as follows:

\begin{align} \delta S^1 = -\frac1{2\pi}  \int_{\tau_{i}}^{\tau_{o}} \rd \tau \int_{\s_{21}}^{\s_{12}} \rd \s \delta \X^{1} \eta \pa_{\s} \S^{1} &+ \frac1{4\pi}  \Big [ \int_{\s_{21}}^{\s_{12}} \rd \s \delta \X (\eta + \om) \pa_{\s}\X \Big ]_{\tau_{i}}^{\tau_{o}} \nonumber \\
\qquad &+ \frac1{4\pi}  \int_{\tau_{i}}^{\tau_{o}} \rd \tau \Big [ \delta \X^1 (2\eta \S^1 - (\eta+\om) \pa_{\tau} \X^1 \Big ] \Big |_{\s = 21}^{\s = 12} . \end{align}

Using the bulk EOM (equation 2.22), we see that the first term vanishes. Thus,

\begin{align} \delta S^1 &= \frac1{4\pi}  \Big [ \int_{\s_{21}}^{\s_{12}} \rd \s \delta \X (\eta + \om) \pa_{\s}\X \Big ]_{\tau_{i}}^{\tau_{o}} + \frac1{4\pi}  \int_{\tau_{i}}^{\tau_{o}} \rd \tau
\Big [ \delta \X^1 2\eta \S^1\Big ] \Big |_{\s = 21}^{\s = 12} \nonumber \\
\qquad &-  \frac1{4\pi}  \int_{\tau_{i}}^{\tau_{o}} \rd \tau
\Big [ \delta \X^1(\eta+\om) \pa_{\tau} \X^1 \Big ] \Big |_{\s = 21}^{\s = 12} \nonumber \\
\qquad &= I + II + III. 
\end{align}

Introducing the notation $\bar{x}_1 =   \int_{0}^{\tau} \S^1$, we calculate the 2nd term of equation (4.13) as

\begin{align}
II &=  \frac1{4\pi}  \int_{\tau_{i}}^{\tau_{o}} \rd \tau \delta (\X^1_{12} - \X^{1}_{21}) 2 \eta \pa_{\tau} \bar{x}_{1}, \nonumber \\
\qquad &=  \frac1{4\pi}  \int_{\tau_{i}}^{\tau_{o}} \rd \tau \delta L_{1} 2 \eta \pa_{\tau} \bar{x}_{1}. 
\end{align}

The variation of the boundary term, $\delta s^1$, is a total variation term which will drop out of the symplectic current calculation, thus, we ignore it. Using equations (4.11), (4.13) and (4.14) we write the total action for strip 1 as

\begin{align}
 \delta S_{T}^1 =  \frac1{4\pi} \Big [ \int_{21}^{12} \rd \s \delta \X^1 (\eta+\om) \pa_{\s} \X^1 \Big ] \Big |_{\tau_{i}}^{\tau_{o}} &+  \frac1{4\pi}  \int_{\tau_{i}}^{\tau_{o}} \rd \tau \delta L_{1} 2 \eta \pa_{\tau} \bar{x}_1 \nonumber \\
\qquad &-  \frac1{4\pi}  \int_{\tau_{i}}^{\tau_{o}} \rd \tau 
\Big [\delta \X^{1} (\eta +\om) \pa_{\tau} \X^1 \Big ] \Big |_{\s = 21}^{\s = 12}. 
\end{align} 

\section{Consistency Check for One-Strip}

Now let us check what our expression (4.15) reduces to in the one-strip case. This will provide us with a sanity check. Comparing with equation (2.35), we see that the 1st term in equation (4.15) is already in the required form. Lets ignore the 2nd term for now and evaluate the 3rd term in the one strip case:

\begin{align}
-III &=   \frac1{4\pi}  \int_{\tau_{i}}^{\tau_{o}} \rd \tau
\Big [ \delta \X^1(\eta+\om) \pa_{\tau} \X^1 \Big ] \Big |_{\s = 21}^{\s = 12} \nonumber \\
\qquad &=  \frac1{4\pi}  \int_{\tau_{i}}^{\tau_{o}} \rd \tau
\Big [ \delta \X^1_{12}(\eta+\om) \pa_{\tau} \X^1_{12} -  \delta \X^1_{21}(\eta+\om) \pa_{\tau} \X^1_{21} \Big ]. 
\end{align}

Let us remember that in the one strip case, $\pa_{\tau} \Delta = \pa_{\tau} (\X^1_{12} - \X^1_{21}) = 0$, which implies that

\be  \pa_{\tau} \X^1_{21} =  \pa_{\tau} \X^1_{12}  \equiv \frac1{2}  \pa_{\tau} (\X^1_{12} + \X^1_{21}). \ee

Using this, it is simple enough to proceed from equation (4.16)

\begin{align}
 \frac1{4\pi}  \int_{\tau_{i}}^{\tau_{o}} \rd \tau
\Big [ \delta \X^1(\eta+\om) \pa_{\tau} \X^1 \Big ] \Big |_{\s = 21}^{\s = 12} &= 
\frac1{4\pi}  \int_{\tau_{i}}^{\tau_{o}} \rd \tau
 (\delta \X^1_{12} - \delta \X^1_{21}) (\eta+\om) \pa_{\tau} \X^1_{12}, \nonumber \\
 \qquad  &= 
\frac1{4\pi}  \int_{\tau_{i}}^{\tau_{o}} \rd \tau
 (\delta \X^1_{12} - \delta \X^1_{21}) (\eta+\om)\frac1{2} \pa_{\tau} (\X^1_{12}+\X^1_{21}) , \nonumber \\
 \qquad  &= 
\frac1{8\pi}  \int_{\tau_{i}}^{\tau_{o}} \rd \tau
 \delta L_1 (\eta+\om) \pa_{\tau} (\X^1_{12}+\X^1_{21}) , \nonumber \\
\qquad &= \frac1{8\pi}  \int_{\tau_{i}}^{\tau_{o}} \rd \tau
\Big [ \pa_{\tau} \Big ( \delta L_1 (\eta + \om) (\X^1_{12}+\X^1_{21}) \Big )\nonumber \\
\qquad & \hspace{10pt}   - (\pa_{\tau} \delta L_{1} (\eta + \omega) \X^{1}_{12} + \X^{1}_{21}) \Big ]. 
\end{align}

Now, using that for the one strip case, $\Delta = L_1$, and $\X^1_{21} = \X_0 \hspace{2pt} \text{and} \hspace{2pt} \X^1_{12} = \X_{2\pi}$ we write

\begin{align}
 \frac1{4\pi}  \int_{\tau_{i}}^{\tau_{o}} \rd \tau
\Big [ \delta \X^1(\eta+\om) \pa_{\tau} \X^1 \Big ] \Big |_{\s = 21}^{\s = 12} 
&= \frac1{8\pi}  \int_{\tau_{i}}^{\tau_{o}} \rd \tau
\Big [ \pa_{\tau} \Big ( \delta \Delta (\eta + \om) (\X_0+\X_{2\pi}) \Big ) - \pa_{\tau} \delta \Delta (\eta + \omega)( \X_0 + \X_{2\pi}) \Big ], \nonumber \\
\qquad &= \frac1{8\pi}  \int_{\tau_{i}}^{\tau_{o}} \rd \tau
\pa_{\tau} \Big [ \delta \Delta (\eta + \om) (\X_0+\X_{2\pi}) \Big ], \nonumber \\
\qquad &= \frac1{8\pi}   \Big [ \delta \Delta (\eta + \om) (\X_0+\X_{2\pi}) \Big ] \Big |_{\tau_{i}}^{\tau_{o}} .
\end{align}

which is the same as in the one strip case (equation 2.35). Thus, we see that our expression reduces to the
one-strip case except the term $\frac1{4\pi}  \int_{\tau_{i}}^{\tau_{o}} \rd \tau \delta L_{1} 2 \eta \pa_{\tau} \bar{x}_{1}$ which might give us a hint of non-oneness.

\section{Strip Symplectic Potential}

Let us now proceed to calculate the symplectic potential from the variation of the on-shell action. Starting from equation (4.15)

\begin{align}  \delta S_{T}^1 &=  \frac1{4\pi}  \int_{21}^{12} \rd \s \delta \X^1 (\eta+\om) \pa_{\s} \X^1 \Big ] \Big |_{\tau_{i}}^{\tau_{o}} +  \frac1{4\pi}  \int_{\tau_{i}}^{\tau_{o}} \rd \tau \delta L_{1} 2 \eta \pa_{\tau} \bar{x}_1 -  \frac1{4\pi}  \int_{\tau_{i}}^{\tau_{o}} \rd \tau 
\Big [\delta \X^{1} (\eta +\om) \pa_{\tau} \X^1 \Big ] \Big |_{\s = 21}^{\s = 12} \nonumber \\
\qquad &= I + II + III. 
\end{align}

The first term is already in the desired form of a difference between initial and final time. The last term can be simplified and we now focus on it.

\begin{align}
III &\equiv  - \frac1{4\pi}  \int_{\tau_{i}}^{\tau_{o}} \rd \tau 
\Big [\delta \X^{1} (\eta +\om) \pa_{\tau} \X^1 \Big ] \Big |_{21}^{12}, \nonumber \\
\qquad &= - \frac1{4\pi}  \int_{\tau_{i}}^{\tau_{o}} \rd \tau 
\Big [\delta \X^{1}_{12} (\eta + \om) \pa_{\tau} \X^1_{12} - \delta \X^1_{21} (\eta + \om) \pa_{\tau}\X^1_{21} \Big ], \nonumber \\
\qquad &= - \frac1{4\pi}  \int_{\tau_{i}}^{\tau_{o}} \rd \tau 
\Big [\delta( \X^{1}_{12} - \X^{1}_{21}) (\eta + \om) \pa_{\tau} \X^1_{12} + \delta \X^1_{21} (\eta + \om) \pa_{\tau}(\X^1_{12} - \X^{1}_{21}) \Big ], \nonumber \\
\qquad &= - \frac1{4\pi}  \int_{\tau_{i}}^{\tau_{o}} \rd \tau \Big [ \delta L_{1} (\eta + \om) \pa_{\tau} \X^1_{12} + \delta \X^1_{21} (\eta + \om) \pa_{\tau} L_{1} \Big ], \nonumber \\
\qquad &= - \frac1{4\pi}  \int_{\tau_{i}}^{\tau_{o}} \rd \tau \Big [ \delta L_{1} (\eta + \om) \pa_{\tau} \X^1_{12} + \delta [ \X^1_{21} (\eta + \om) \pa_{\tau} L_{1}  ] 
- \pa_{\tau}[\X^1_{21} (\eta + \om) \delta L_{1} ] + \pa_{\tau} \X^1_{21} (\eta + \omega) \delta L_{1} ], \nonumber \\
\qquad &\hat{=} \frac1{4\pi} [\delta L_{1} (\eta - \om) \X^1_{21}] |_{\tau_{i}}^{\tau_{o}} - \frac1{4\pi} \int_{\tau_{i}}^{\tau_{o}} \rd \tau [ \delta L_{1} (\eta + \omega) \pa_{\tau}\X^1_{12} + \delta L_{1} (\eta - \omega) \pa_{\tau}\X^1_{21} ].
\end{align}

In the last line we have neglected the total variation term and denoted the equality up to terms $\delta s$ by a
hatted equality. We can obtain the same result by exchanging the role of $\X^1_{12}$ with $\X^1_{21}$. This gives

\be III \hat{=} \frac1{4\pi} [\delta L_{1} (\eta - \om) \X^1_{12}] |_{\tau_{i}}^{\tau_{o}} - \frac1{4\pi} \int_{\tau_{i}}^{\tau_{o}} \rd \tau [ \delta L_{1} (\eta + \omega) \pa_{\tau}\X^1_{21} + \delta L_{1} (\eta - \omega) \pa_{\tau}\X^1_{12} ]. \ee

By averaging these two contributions we obtain

\be III \hat{=} \frac1{4\pi} \Big [ \delta L_{1} (\eta - \om) \frac1{2} (\X^1_{12} + \X^1_{21}) \Big ] \Big |_{\tau_{i}}^{\tau_{o}}  - \frac1{4\pi} \int_{\tau_{i}}^{\tau_{o}} \rd \tau [\delta L_{1} \eta \pa_{\tau} (\X^1_{21} + \X^1_{12}) ]. \ee

In summary we find that the strip action is the sum

\begin{align} \delta S_{T}^{1} &=  \frac1{4\pi} \Big [ \int_{21}^{12} \rd \s [ \delta \X^1 (\eta + \omega) \pa_{\s} \X^1 ] +  \frac1{4\pi}  [ \delta L_1 (\eta -\om)  \frac1{2} (\X^1_{12} + \X^1_{21})  ] \Big ]_{\tau_{i}}^{\tau_{o}} \nonumber \\
\qquad & - \frac1{4\pi} \int_{\tau_{i}}^{\tau_{o}}  \rd \tau [ \delta L \eta \pa_{\tau}(2 \bar{x}_{1} - (\X_{21}^1 + \X^1_{12} ) ) ] . 
\end{align}
 
Looking at the above expression, we observe that the first two terms are in the form we desire but the last term is not. However, it also gives us a hint. The last term involves the integration constant $\bar{x}_1$ which still has to be fixed. It is natural then to demand that its value makes the last term vanish. Thus,

\be \bar{x}_1 = \frac1{2}(\X_{21}^{1} + \X^1_{12}). \ee

It shows that $\S^1 = \pa_{\tau} \bar{x}_1$ is the velocity of the strip midpoint. Given this choice we can now write the symplectic potential up to total variation as

\be \Theta^1 \hat{=} \frac1{4\pi}  \int_{21}^{12} \rd \s [ \delta \X^1 (\eta + \omega) \pa_{\s} \X^1 ] + \frac1{4\pi} \delta L_1 (\eta - \omega) \bar{x}_1 . \ee

From this expression we see that the strip extension $L_1$ is conjugate to the midpoint $\bar{x}_1$. This extra data represent the strip new degrees of freedom. The symplectic structure $\Omega$ can be calculated as:

\be \Omega = \delta \Theta. \ee

Thus, from equation (4.26), we can calculate, the strip symplectic structure for Strip 1 as:

\[
\boxed { \Omega^1 = \frac1{4\pi}  \int_{21}^{12} \rd \s \Big [ \delta \X^1 \eta \pa_{\s} \delta \X^1 + \delta L_1 \eta \delta \bar{x}_1 \Big ]. } \]

Looking at the above expression, we see that the two form $\omega$ has completely dropped out of the expression for the symplectic structure, which is consistent with our demand that it must be closed in $\omega$. This means that the $N$ strip symplectic structure as will derive later, is also independent of $\omega$. Thus, the strips can be glued back together and we have succeeded in recovering the closed strip worldsheet from the strip picture!!

\section{N Nakamura Strips}

\begin{figure}[ht]
	\centering
	\includegraphics[scale=0.5]{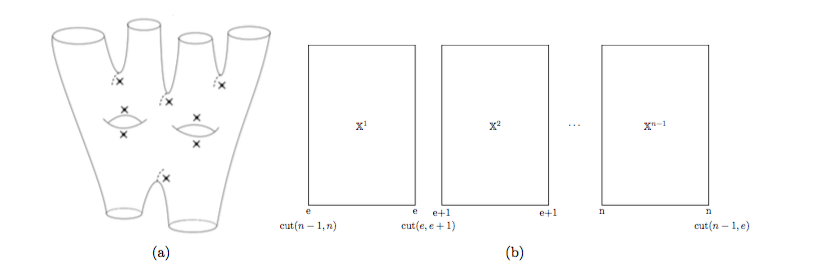}\label{fig4}
	\caption{(a) A general closed string interaction pants diagram, (b) General-Nakamura decomposition}
	\end{figure}

Given a general surface, we would like to decompose it back to a collection of flat nakamura strips. The number of Nakamura strips might be $N$, where $N$ is arbitrary. Take for example, a pants diagram with a single interaction point as in Figure 3.1. In that case, it suffices to decompose the worldsheet in terms of two strips as in Figure 3.2. But what if we have $N$ interaction points where $N > 2$, for example, Figure 4.1(a)? In that case, we need to generalize the analysis we carried out in the previous chapter for two strips. In this section, we will write the equations of motion for the bulk and boundary for the gluing of $N$ Nakamura strips. We will also deviate from the specific notation we used in the previous chapter for ease of generalization in the following manner: we will rename the cuts to be at $\s_{e,e+1}$ where $(e,e+1)$ denote the cut between strips $e$ and $e+1$ (Figure 4.1b).

In the case of two cuts, we look at a strip initially cut at $(0,2\pi)$ and then cut again at $\s = \s_{e,e+1}$. As before, we impose $\delta \X(\tau_0,\s) = \delta \X(\tau_i,\s) = 0$, and thus, the onshell variation of the action can be written as before (the matrix indices are implicit for keeping it simple) :

\begin{small}

\be  \delta S \hat{=} \frac1{4\pi} \int_{\tau_{i}}^{\tau_{o}} \rd \tau \Big [ \delta \X_{e} (2\eta \S_e - (\eta + \om) \pa_{\tau} \X_e ) \Big ] \Big |_{\s_{e+1,e}}^{\s_{e,e+1}} + \frac1{4\pi} \int_{\tau_{i}}^{\tau_{o}} \rd \tau \Big [ \delta \X_{e+1} (2\eta \S_{e+1} - (\eta + \om) \pa_{\tau} \X_{e+1} ) \Big ] \Big |_{\s_{e,e+1}}^{\s_{e+1,e}}. \ee

\end{small}

Let us recall that for the moment, we restrict our analysis to the case where $(\eta,\om,H)$ are all constant and that each cut e belongs to two strips $S_e$ and $S_{e+1}$. The continuity of the action demands that across a cut located at $\s_{e,e+1}$, we have $\pa_{\s}\X_{e}(\s_{e,e+1}) = \pa_{\s}\X_{e+1}(\s_{e,e+1}) $. The discontinuities in the choice of coordinates are encoded into the edge monodromies which encode the translation involved with the change
of coordinates $\X_e  \rightarrow \X_{e+1}$

\be \Delta_{e,e+1} \equiv \X_{e+1} (\s) - \X_{e} (\s) = \int_{\s_e}^{\s_{e+1}} \eta . \ee

As in the previous chapter, we use the bulk equations of motion to see that, at each cut, $\s = \s_e$, we have

\be \pa_{\tau} \Delta_{e,e+1} = \S_{e+1} - \S_{e}. \ee

Following our analysis as we have in the two strip case, we find that for $N$ strips, the total variation of the action can be written as a sum over strips (sum over $e$'s) as follows:

\be  \delta S_{T} = \frac1{4\pi} \sum_{e=1}^{N-1} \int \rd \tau \Big ( \delta \Delta_{e,e+1} [2\eta \S_{e+1} - (\eta + \om) \pa_{\tau} \X_{e,e+1}^{e+1} ] +  \delta \X_{e,e+1}^{e+1} [ (\eta - \om) \pa_{\tau} \Delta_{e,e+1} ] \Big ) . \ee

Demanding that this variation vanishes, we get the following general boundary equations of motion:

\be \sum_{e} (\eta - \om) \pa_{\tau} \Delta_{e,e+1} = 0 \ee

\be  \sum_{e} 2\eta \S_{e+1} - (\eta + \om) \pa_{\tau} \X_{e,e+1}^{e+1} = 0, \ee

which are consistent with the two strip and one strip case.

\section{$N$ -Strip Symplectic Potential}

Looking at the symplectic potential, equation (4.26), for strip 1 in the 2-strip case, we can generalize it to write the $N$ -strip sympectic potential as a summation over strips as follows:

\be \Theta \equiv \sum_{e}^{n-1}\Theta^{e} \hat{=}  \sum_{e}^{n-1} \frac1{4\pi} \int_{n,e}^{e,e+1} \rd \s [\delta \X_{e} (\eta+\om) \pa_{\s} \X_{e}] + \frac1{4\pi} \delta L_{e} (\eta - \om) \bar{x}_{e}. \ee

\section{$N$ - Strip Symplectic Structure}

Looking at equation (4.28), we also know how to write the $N$ strip symplectic structure as a sum over strips as follows:

\be \Omega = \sum_{e} \Omega^e = \frac1{4\pi} \int_{n,e}^{e,e+1} \rd \s [\delta \X_{e} \eta \pa_{\s} \delta \X_e + \delta L_e \eta \delta \bar{x}_e ], \ee

where, $\bar{x}_e$ is the mid point of a generic strip $e$. Looking at the above expression, we see that it is free of $\om$ which implies that the strips can be safely glued back together. It is indeed consistent with our interpretation that the symplectic flux must be continuous across the boundaries. Hurray!

\section{Discussions}

Looking at the expression for the $N$-strip symplectic potential, equation (4.35) in the previous section, we see that for each strip, we have 2 strip variables namely, $L_e$ and $\bar{x}_e$. The amplitude as well as the symplectic current can be written as a function of these two strip variables. Looking at the last term in the expression for the symplectic current, we see that these strip variables look like conjugates of each other, where $\bar{x}_e$ denotes the mid-point of each strip, and $L_e$, the strip extension which is the conjugate momentum to the mid-point of the strip, $e$. Following from section (4.3), we discover that for each strip, $\S_e = \pa_{\tau} \bar{x}_e$ is the velocity of the mid-point of the strip. Moreover, since the symplectic structure is closed in $\om$, we have recovered the closed string worldsheet successfully! This is indeed cause for celebration!

\begin{savequote}[45mm]
It is entirely possible that beyond the perception of our senses, worlds are hidden of which we are unaware.
\qauthor{Albert Einstein}
\end{savequote}

\chapter{Conclusion}

Hoping that we have made sense of some more of the ``storm clouds in theoretical physics" [5] and gotten closer to dealing with the question of space-time, we summarize and conclude the essay here.

To start with, this essay reviewed the salient features of meta-string theory and the motivation behind proposed modifications in string theory. The departure to a phase-space-target space has interesting ramifications. As discussed before, we see that it leads to T-duality getting manifest in the target space. It also provides a natural scope of curving the momentum space thus leading to a hope of unifying GR and QM in a natural way.

We then moved on to deal with the study of the Nakamura-strips in more detail. Meta-string theory removes the assumption of the background space-time in which a string propagates. This naturally leads to breaking the periodicity condition of $\X$ and introduces a monodromy in $\X$, where $\X$ is the phase space variable. But this no more naturally allows for a closed string condition! We reviewed the gluing condition for an open strip as analysed in [2] where the closed worldsheet has been recovered by demanding that the symplectic current be conserved at the boundary of the strips or ``cuts''.

Now, looking at Figures 3.2 and 4.1(a), it is very apparent that we would have to glue more than one strip together in weird ways in order to form a general closed string interaction picture. In chapter 3, we take to this job with gusto. We find the expression for the closed meta-string variation of the action as a sum over two Nakamura-strips and derive the bulk and boundary equations of motion. The bulk equations of motion can be seen to decouple for each of the strips which is very natural. We then generalize it to an arbitrary number of strips.

The boundary equation of motion on the other hand, derived by demanding that the variation of the action vanishes at the boundary, imply that $(\eta + \om)\pa_{\tau} L_e = 0$ for both boundaries of any generic strip, $e$. This is the Dirichlet boundary condition term and we see the natural evolution of D-branes in the meta-string formalism. In the process, we cannot glue these strips back together and we have derived the conditions for the open meta-strips ending on meta- D branes on both sides. Moreover, we see that $ \pa_{\tau} L_e$ is in the kernel of $(\eta + \om)$ which is the projector on to the space-time lagrangian sub-manifold implying that the meta-D branes exist in the space-time lagrangian sub-manifold of the phase-space.

Chapters 4 proceeds to look at the symplectic structure of the closed strip gluing condition. Demanding the action to be invariant of the position of the cuts as well as orientation independent, we recover a very simple boundary term to be added for each strip. We then proceed to vary the action and further recover the symplectic form for each strip. We are pleased to find that the symplectic form for a N -strip case can be written simply as a sum over strips for a general $e^{th}$ strip. We compare the symplectic form we derived for the general strip case with the one-strip case and see that there is an extra term which encodes the non-oneness. Upon closer inspection, we see that this term seems to be similar to the symplectic form for a particle. Positing $\bar{x}$ as the mid-point of a given strip, we see that we can interpret $L$ as the momentum which is conjugate to the strip mid-point. It implies that the variation of the strip separation governs the amount of symplectic flux flowing through the boundaries of each strip, though at this point, this interpretation is strictly conjectural.

\section{Future Directions}

\begin{figure}[ht]
	\centering
	\includegraphics[scale=0.5]{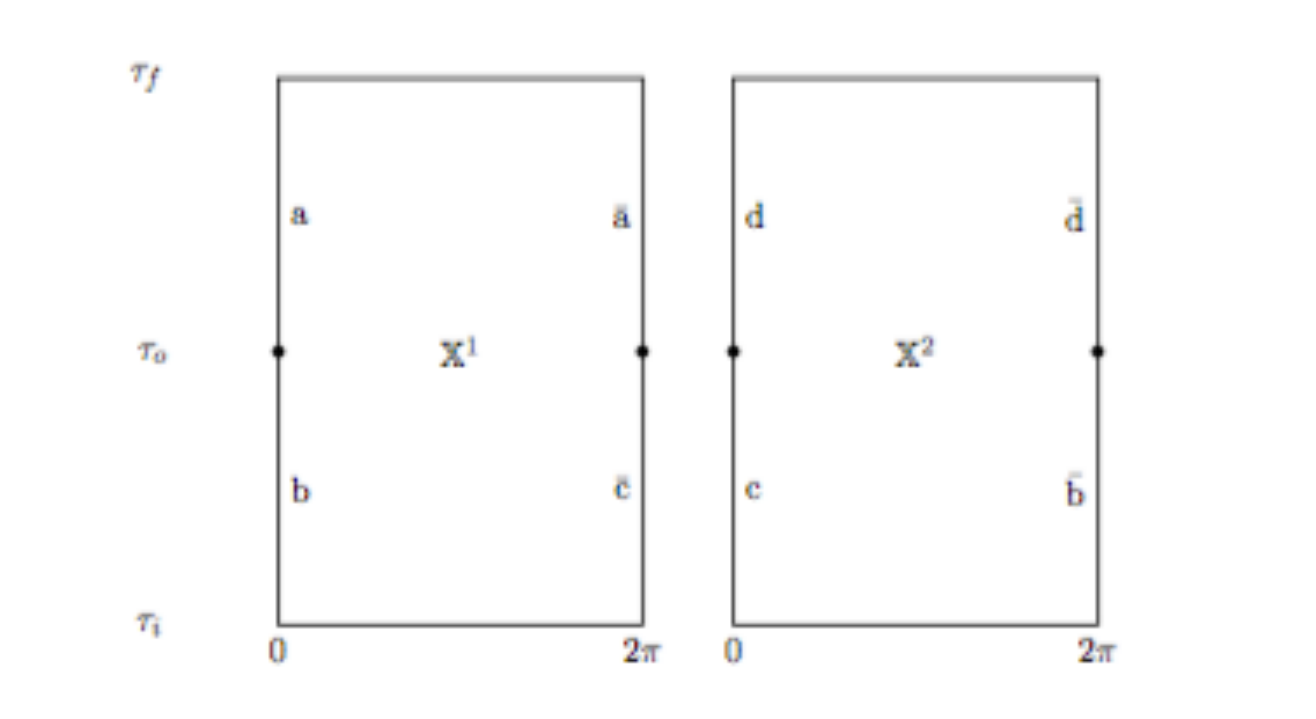}\label{fig5}
	\caption{Two-Strip Interaction picture for the Pants diagram 3.1.}
	\end{figure}

Looking at the strip picture, we are reminded of the string-bit proposal by Klebanov and Susskind. Here, even though we are dealing with the semi-classical picture, the strips look like the classical analogs of the string bits. It would be the next step to quantize the theory and compare our results and see what it predicts regarding the back ground space-time. Looking at the results of [3] and [4], where the string bits experience a discrete back-ground space-time in the continuum limit, it would be interesting to study the results for the background phase-space in the case of the meta-string.
The meta-string formalism also gives a vague hint of being able to unify the Chan-Paton factors of the matrix models of string theory in a natural way. If we look at equation (4.10), we see that we can write the total amplitude $\Am$ as a product of strip amplitudes, $\Am_e$:

\be \Am = \prod_{e} \int_{\tau_{i}}^{\tau_{f}} \rd \tau \int \rd L_{e} \rd \X_{e} \Am_{e} \ee

where

 \[
 \boxed { \Am_{e}(L_{e}, \bar{x}_{e}) = \int_{C_e} [D\X] e^{i S_{T}^{e}(\X_{e})}  }
 \]  

In the above equation

\be L_{e} = \X_{e,e+1}^{e} - \X_{e-1,e}^{e} \ee

is the generic strip separation vector and

\be \bar{x}_{e} = \frac{\X_{e,e+1}^{e} + \X_{e-1,e}^{e}}{2} \ee

is the generic strip midpoint for any strip $e$. The integration over the boundary is defined as

\be \int_{C_e} = \frac{1}{(\s_{e,e+1}- \s_{e-1,e})} \int_{e-1,e}^{e,e+1} \rd \s . \ee

If we might make a guess, we might say that the formalism provides a scope for naturally finding out the Chan-Paton factors and to write the interaction amplitude as a product of strip matrices which depend on the strip variables $L_e$ and $\bar{x}_e$. Looking at the form of the symplectic structure, we posit that $L_e$ and $\bar{x}_e$ are the  \textbf{canonically conjugate variables} associated with each strip. This opens the door to canonically quantizing the strip picture. There is also a relation between the fixing of the strip mid-point $\bar{x}_{e}$ to time translation symmetry in meta-string theory, which has not yet been probed in this book. Moreover, this is interesting because this shows us the path to a very simple way of writing any arbitrary string interaction amplitudes just as a permutation of strip-gluing. For example, let us consider the following two strip interaction diagram in Figure 5.1. The amplitude can be divided into two parts in $\tau$ (see Figure 5.1),

\be 1: \tau_{i}\hspace{2pt} \text{to} \hspace{2pt}\tau_{o}, \ee
\be 2: \tau_{o} \hspace{2pt}\text{to}\hspace{2pt} \tau_{f}. \ee

To create the pants in Figure 3.1, we can write the gluing condition in tabular format as shown below in Figure 5.2. Written in terms of the strip variables, $\X^1$ and $\X^2$, the matrix would look like Figure 5.3. As 

\begin{figure}[ht]
	\centering
	\includegraphics[scale=0.5]{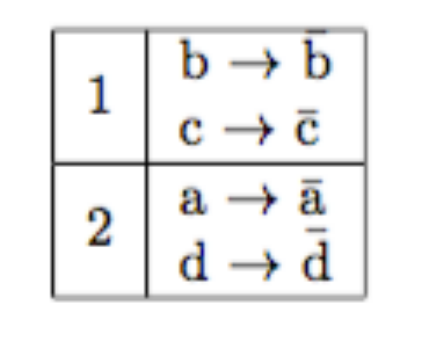}\label{fig6}
	\caption{Gluing condition for a single interaction diagram such as Figure 3.1.}
	\end{figure}
	
\begin{figure}[ht]
	\centering
	\includegraphics[scale=0.5]{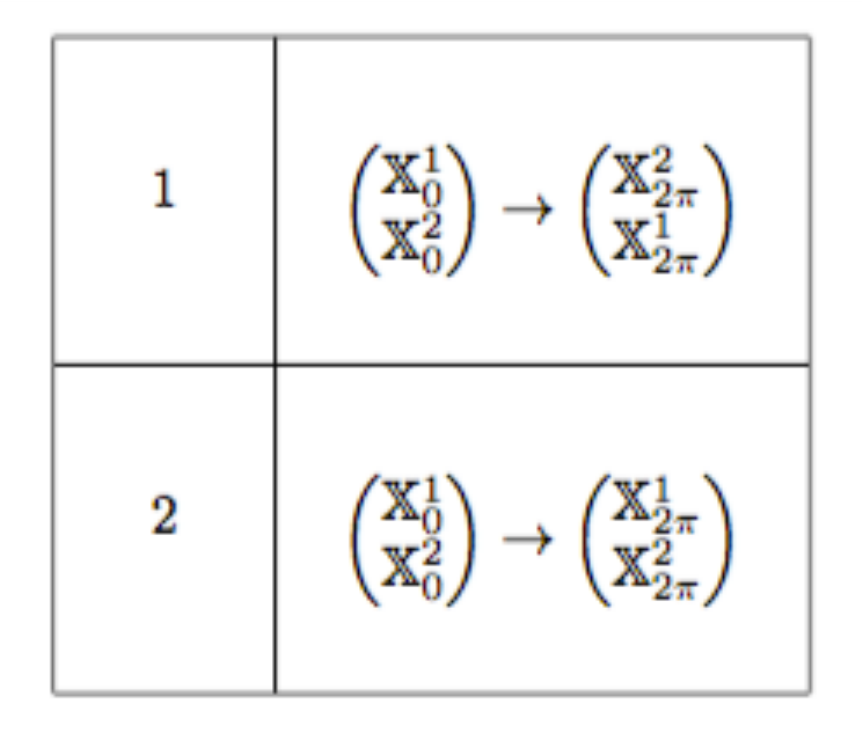}\label{fig7}
	\caption{Gluing condition for a single interaction diagram in terms of 2-strip variables.}
	\end{figure}

we can see from Figure 5.3, going from $1 \rightarrow 2$ is just a permutation of the strip variables $\X^1_{2\pi}$ and $\X^2_{2\pi}$ (flipping the second vector), which effectively changes the gluing condition according to the evolution of the interaction in $\tau$. This tells us that interaction points can be translated as the insertion of a permutation operator $\hat{\mathbb{O}}$ in the expression for the total amplitude, equation (5.1) as follows:

\be \Am = \prod_{e}  \Big [  \int_{\tau_{i}}^{\tau_{o}} \rd \tau \rd L_e \rd \X_e \Am_e +   \int_{\tau_{o}}^{\tau_{f}}  \rd \tau \rd L_e \rd \X_e  \hat{\mathbb{O}}  \Am_e \Big ]. \ee

The above expression is remarkably simple and would allow us to calculate the amplitude for all possible kind of string interaction pictures such as Figure 4.1(a) with great ease given any given kind of string theory. Currently, there exists no simple way to carry this out! Surprisingly, this also hints that interactions in strings can be treated non-perturbatively, even though it has always been assumed that interactions are inherently perturbative.

\chapter*{Bibliography}
\addcontentsline{toc}{chapter}{Bibliography}

\begin{itemize}

\item[1] V. F. Rovelli, Carlo, Covariant Loop Quantum Gravity. Cambridge University Press, 2014.

\item[2] L. Freidel, R. G. Leigh, and D. Minic, Metastring Theory and Modular Space-time, arXiv:1502.0800.

\item[3] I. Klebanov and L. Susskind, Continuum strings from discrete field theories, Nuclear Physics B 309 (1988), no. 1 175 – 187.

\item[4] M. Karliner, I. Klebanov, and L. Ssusskind, Size and shape of strings, International Journal of Modern Physics A 03 (1988), no. 08 1981–1996.

\item[5] E. Witten, Reflections on the fate of space-time, Physics Today 49N4 (1996) 24–30.

\end{itemize}

\chapter*{}

\end{document}